\documentclass[preprint, journal]{vgtc}                

\ifpdf
  \pdfoutput=1\relax                   
  \usepackage{graphicx}                
  \DeclareGraphicsExtensions{.pdf,.png,.jpg,.jpeg} 
\else
  \usepackage{graphicx}                
  \DeclareGraphicsExtensions{.eps}     
\fi%

\graphicspath{{figures/}{pictures/}{images/}{./}} 

\usepackage{microtype}                 
\PassOptionsToPackage{warn}{textcomp}  
\usepackage{textcomp}                  
\usepackage{mathptmx}                  
\usepackage{times}                     
\usepackage{cite}                      
\usepackage{tabu}                      
\usepackage{booktabs}                  
\usepackage{wrapfig}
\usepackage{subfig}
\usepackage{enumitem}
\usepackage[english,ruled,vlined]{algorithm2e}
\usepackage{amsmath, amsfonts, amssymb}

\vgtccategory{Research}
\vgtcpapertype{application/design study}

\title{Preserving Command Line Workflow for a Package Management System using
ASCII DAG Visualization}
  
\author{
  Katherine E. Isaacs and Todd Gamblin
}
\authorfooter{
\item
  Katherine E. Isaacs is with the University of Arizona\\
  E-mail: kisaacs@cs.arizona.edu.
\item
  Todd Gamblin is with Lawrence Livermore National Laboratory\\
  E-mail: tgamblin@llnl.gov
}

\abstract{
  Package managers provide ease of access to applications by removing the
time-consuming and sometimes completely prohibitive barrier of successfully
building, installing, and maintaining the software for a system. A package
dependency contains dependencies between all packages required to build and
run the target software. Package management system developers, package
maintainers, and users may consult the dependency graph when a simple listing
is insufficient for their analyses. However, users working in a remote command
line environment must disrupt their workflow to visualize dependency graphs in
graphical programs, possibly needing to move files between devices or incur
forwarding lag. Such is the case for users of Spack, an open source package
management system originally developed to ease the complex builds required by
supercomputing environments. To preserve the command line workflow of Spack,
we develop an interactive ASCII visualization for its dependency graphs.
Through interviews with Spack maintainers, we identify user goals and
corresponding visual tasks for dependency graphs. We evaluate the use of our
visualization through a command line-centered study, comparing it to the
system's two existing approaches. We observe that despite the limitations of
the ASCII representation, our visualization is preferred by participants when
approached from a command line interface workflow. 

} 


  \teaser{
  \centering
 \includegraphics[width=16.3cm]{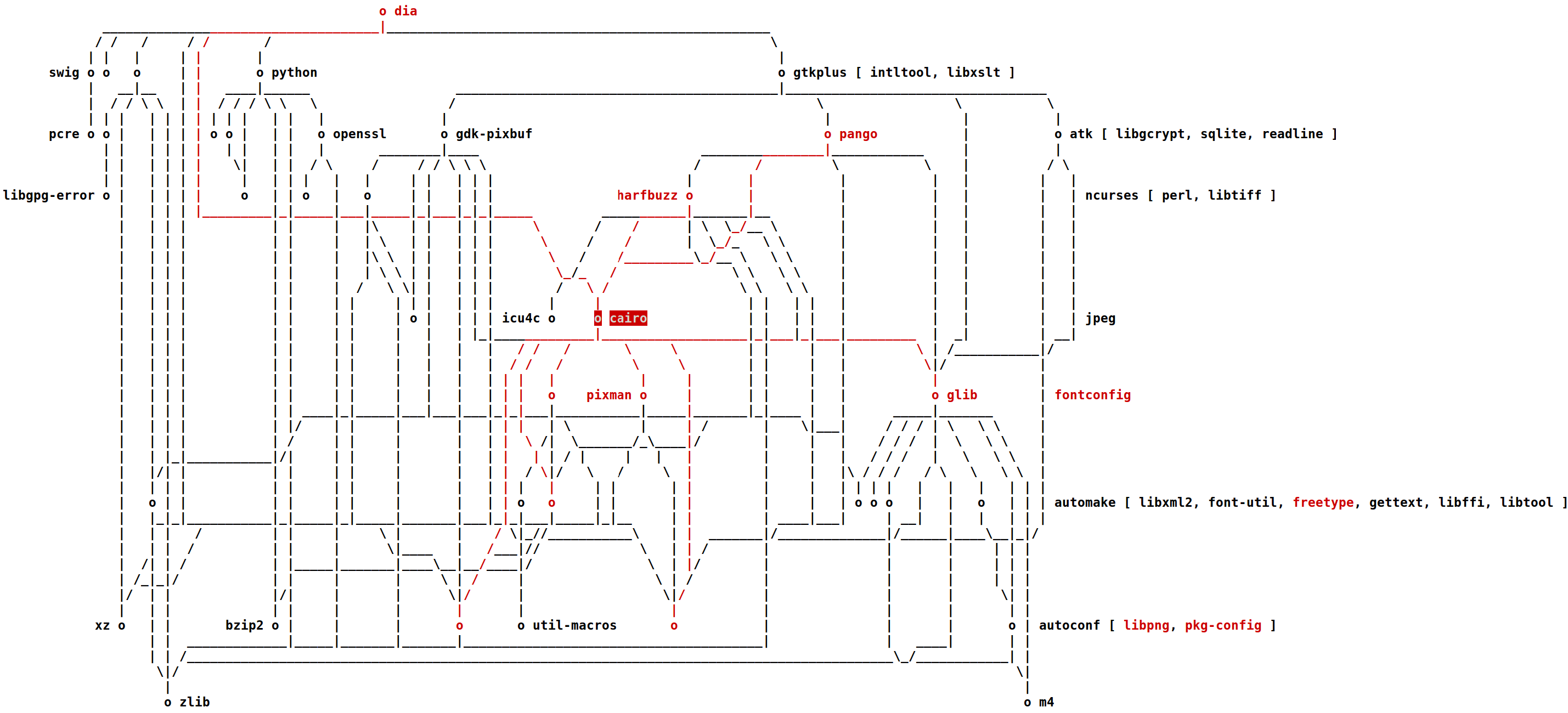}
\caption{ASCII depiction of the package dependency graph of \texttt{dia} (40
nodes), as specified in the Spack package management tool. The dependency
\texttt{cairo} and its direct neighbors have been interactively highlighted.}
\label{fig:teaser}
}

\vgtcinsertpkg

\begin{document}


\firstsection{Introduction}

\maketitle

\label{sec:intro}
Building and executing software can be a complicated and frustrating process
due to complex requirements in terms of dependencies, their versions and
install locations, and how they were compiled. Package management systems,
such as Homebrew~\cite{homebrew} for OSX systems and APT~\cite{apt} for Ubuntu
systems, have been developed to handle these complexities automatically for
most users. Use of these package management systems can thus expand the
usability of software by greatly decreasing the burden on users attempting to
get software to run.

For users of a particular software application to benefit, someone must first
{\em package} the software for the system. This is often done by the engineer
or scientist who developed the application. Furthermore, the developers of the
package management system itself expand and maintain functionality as build
systems and architectures evolve. Additionally, power users with unique needs
may want to specify additional requirements when building the software. In all
of these scenarios, understanding the dependencies of a particular software
package can be beneficial. While a list or tree of packages is sufficient for
some tasks, sometimes consulting the full set of package relationships is of
interest. Generally the relationships among a package and all of its direct
and indirect dependencies form a directed acyclic graph (DAG). As these graphs
generally are limited to tens of nodes and developers are particularly
interested in the dependency paths, the graph is frequently represented as a
node-link diagram.

\begin{figure*}[bth]
    \centering
  \subfloat[\texttt{radare2}\label{fig:radare2:r2}]{
    \includegraphics[width=0.57\textwidth]{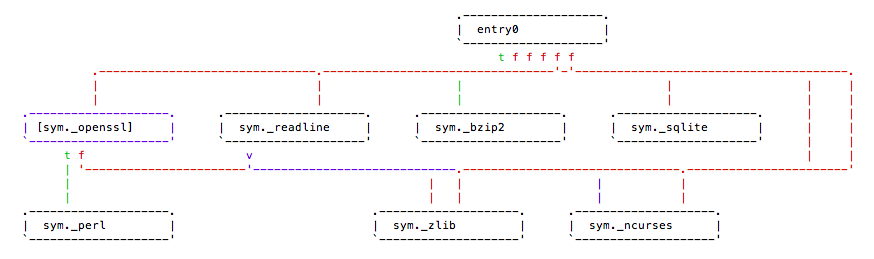}
}
\hspace{2ex}
  \subfloat[\texttt{graphterm}\label{fig:radare2:term}]{
    \includegraphics[width=0.38\textwidth]{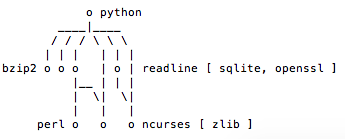}
}
    \caption{Package dependency graph of python rendered through (a)
\texttt{radare2}'s call graph feature, and (b) our approach.}
    \vspace{-0.05in}
\label{fig:radare2}
\end{figure*}
Many package management systems are command line tools.  We focus on one such
open source tool, Spack~\cite{Gamblin2015Spack, SpackGithub}, which was
developed to meet the needs of supercomputing applications. In this context,
developers may only be able to access the supercomputer or other remote system
via a terminal interface.  The most expedient method of visualizing a small
DAG often requires several extra steps in transferring the data to a local
machine and running a separate program instead of the terminal in which the
user was working. This step can represent a significant overhead in terms of
the entire analysis needed. These difficulties with accessing plotted data
from command line workflows have led to the creation of ASCII charting support
in tools such as gnuplot~\cite{gnuplot} and Octave~\cite{Eaton2014}.  To
better match the workflow of command line-focused users when viewing a DAG, we
propose bringing interactive DAG visualization to the terminal. Our goal is to
``meet command line users where they are'' when possible.

We developed an interactive, terminal-based DAG visualization to support the
analysis of package dependency graphs. The limitations of the terminal pose
many challenges. To ensure portability, marks are limited to ASCII characters
that cannot be overlaid. Available colors are limited as well. Relatively few
marks can be shown simultaneously due to the row and column size of the
terminal. With these limitations in mind, we convert a graphical layered graph
layout to ASCII. Search and highlight capabilities are available via keyboard
commands. We conducted a study comparing the efficacy of the command line
workflow with our visualization to the workflows using the existing
GraphViz-rendered \texttt{dot}~\cite{GraphViz,Gansner1993dot} layout and
static ASCII representations. We find that despite the limitations of the
ASCII visualization, it can be effective in a command line workflow and our
study participants preferred it to the existing workflows. 

In summary, our contributions are:
\vspace{-0.02in}
\begin{enumerate}
  \item A domain and task characterization for package dependency graphs
    (Sec.~\ref{sec:background}),
  \vspace{-0.1in}
  \item an interactive visualization of directed acyclic graphs using only the characters 
    $\{$ \texttt{|}, \texttt{\_}, \texttt{/}, \texttt{$\backslash$},
    \texttt{X}, \texttt{o} $\}$ (Sec.~\ref{sec:ascii}), 
  \vspace{-0.1in}
  \item a study comparing the use of our interactive terminal-based
    DAG visualization in a command line workflow to that of 
    the existing graphical and ASCII solutions
    (Sec.~\ref{sec:study} and Sec.~\ref{sec:analysis}), and
  \vspace{-0.1in}
\item a survey of dependency visualizations provided by Github projects
  (Sec.~\ref{sec:related} and appendices).
\end{enumerate}

%

\section{Related Work}
\label{sec:related}

Several graphical visualizations, such as SHriMP Views~\cite{Storey1996},
ExTraVis~\cite{Cornelissen2007}, SolidSX~\cite{Reniers2011}, the layouts
available through Roassal's GRAPH feature~\cite{Bergel2014}, and the work of
Noack and Lewerentz~\cite{Noack2005}, have been proposed for software
dependencies, but most focus on the software creation and maintenance
aspect---how the dependencies relate to developing the software---rather than
the packaging and building process. One difference is that software module
dependencies frequently exhibit a containment structure not present in our
package dependency data but emphasized by the existing visualizations.
Another is that some visualizations make design decisions to accommodate scale
or temporal behavior that our package dependency data does not exhibit. Kula
et al.~\cite{Kula2014} developed a system to show changes in a software's
library dependencies over time, but did not directly show the relationship
between the libraries, only with the target software. The visualization was
for maintainers of the software, not people trying to install the software or
develop installation management tools. For a survey of software dependency
visualization, see Caserta and Zendra~\cite{Caserta2011}.

\vspace{1ex}

\noindent\textbf{Dependency Visualization in Github Repositories}.  Many open
source software tools offer some form of dependency visualization.  We
surveyed Github repositories retrieved in a search for ``visualize
dependencies''~\footnote{See appendix for methodology, charts
describing all views, tools, and formats observed, and see supplemental
materials for a complete list of
projects surveyed.} to determine what visualizations are available and which
are commonly employed by open source developers. We analyzed a broad set of projects including many types of
dependency visualizations in computing, not only those used by package
managers. Some projects offer multiple visualization features.
Of the projects with visualization features (224 total of 483 surveyed),
57.6\% use some form of hierarchical layout node-link diagram. Other common
representations were force-directed node-link diagrams (23.2\%) and indented
or node-link trees (15.2\%). We also found a few instances of other
visualizations showing tree structure (sunburst, treemap, flamegraph,
CodeCity), as well as a Sankey diagram, an arc diagram, and a chord diagram.
Force-directed layouts and chord diagrams do not capture the DAG structure of
package dependency graphs. Trees are useful for some package manager tasks
(\ref{sec:background}), but our work is motivated by supporting graph
topology tasks (\ref{sec:taskabstraction}) that are not well supported by the
existing indented trees.

Of the projects in our Github survey, 48.2\% use GraphViz in some form, with
the most common usage (65.7\%, or 31.7\% of all projects with visualization
features) being to output a \texttt{dot} format file with documentation
suggesting rendering with GraphViz's \texttt{dot} algorithm. The next most
common tool was d3js, in use by 21.0\% of the projects. ASCII indented trees
were found in 9.8\% of the projects. Aside from one ASCII graph tool
(``\texttt{ascii-graph}'') described below and a single view based on \texttt{git}
similar to the one described in \ref{sec:git}, only the ASCII indented
tree and the non-rendered graph description files (e.g., \texttt{dot},
\texttt{graphML}, \texttt{GEXF}) can be viewed from the command
line without a separate application, but these do not explicitly show the
topological graph features of interest.

\vspace{1ex}

\noindent\textbf{Graph Drawing in ASCII}.
ASCII flow chart tools were initially created to aid print documentation of
computer programs. In these depictions, nodes take the form of outlined boxes
with labels inset. FlowCharter~\cite{Haibt1959} broke programs into chunks of
six to seven nodes at various levels of abstraction in creating flow charts.
Knuth~\cite{Knuth1963} printed each box in sequence vertically and routed
orthogonal edges along the side. 

There are many tools to assist manual creation of ASCII node-link diagrams for
electronic documentation, such as Emacs Artist~\cite{EmacsArtist}, ASCII
Flow~\cite{ASCIIFlow2}, and AsciiO~\cite{ASCIIO}, but fewer that do automatic
layout. \texttt{Graph::Easy}~\cite{GraphEasy}, \texttt{vijual}~\cite{vijual},
and \texttt{ascii-graphs}~\cite{ascii-graphs} are general tools for
automatically producing static ASCII layouts with similar node styles to
FlowCharter and the diagrams of Knuth. As one of our design goals is to
represent the graph topology compactly, we did not want to use a boxed node
style. Furthermore, of the three, only \texttt{ascii-graphs} offers a
hierarchical layout that would match the character of our package dependency graphs.

The software reverse engineering framework \texttt{radare2}~\cite{radare2} can
display code branching graphs and call graphs in ASCII. Similar to our
approach, it is based on a layered graph algorithm and provides interactivity.
It was the only other ASCII graph representation we found that did so.
However, we determined the depictions and interactivity were not suitable for the
character of our package dependency graphs or the tasks of interest on them.
Fig.~\ref{fig:radare2:r2} shows the \texttt{radare2} call graph of a program we
wrote to mimic the dependency graph of python as packaged by Spack
(Sec.~\ref{sec:background}).  Fig.~\ref{fig:radare2:term} shows our depiction of
the same graph for comparison. 

General image to ASCII converters~\cite{OGrady2008ASCIIConversion,
Markus2015ASCIIArt, Takeuchi2013ASCIIArt} focus on
emulating tone differences across an image and thus are a poor fit for
node-link diagrams. Xu et al.~\cite{Xu2010} present an algorithm for
converting vector line art to ASCII, but their method requires several minutes
to generate an image, which is unacceptable in our workflow. Furthermore, none
of the general ASCII conversion methods preserve the structural meaning of the
elements, and thus would require a post-processing step to identify the
vertices and edges to support interactivity.

In Table~\ref{tab:features}, we summarize the most suitable approaches and their
support for desired features in representing package dependency graphs. 

\vspace{1ex}

\noindent\textbf{General Graphical Layout Approaches}.  Many layered layout
algorithms~\cite{Sugiyama} and orthogonal layout
algorithms~\cite{Tamassia1987} use graphical marks. For
a more in depth discussion of these algorithms, see
Tamassia~\cite{GraphDrawingHandbook}. We select a layered layout for
conversion to ASCII as described in Sec.~\ref{sec:ascii}, finding the structure
matches both the tasks and expectations of our users.  For example, layered
layouts make clear the direction of dependencies by position. Flow direction
is not a priority in many orthogonal layouts.  The optimally compact grid
layout of Yoghourdjian et al.~\cite{Yoghourdjian2016} balances directional
flow requirements with other orthogonal aesthetic criteria and typographical
content constraints by using multiple directions for flow. We chose to not
relax flow direction as they do in order to preserve user expectations (i.e.,
vertical dependency direction), as an evaluation of graph drawing aesthetic
criteria for UML diagram comprehension~\cite{Purchase2001UML} suggests domain
semantics play an important role.  Dwyer et
al.~\cite{Dwyer2013EdgeCompression} use grouping techniques to decrease edge
clutter in directed graphs, making them more compact. However, these
techniques use marks to denote group containment and we are constrained in the
number of marks we can legibly fit in a terminal window when using ASCII.

\begin{table}
  \caption{Summary of Support for Desired Features}
  \label{tab:features}
  \scriptsize
  \centering
  \begin{tabu}{
      *{6}{c}
    }
  \toprule
    Feature & graphterm & dot & git log & ascii-graphs & radare \\
            &           & PDFs & graph  &              & \\
  \midrule
    displays in & Yes & No & Yes & Yes & Yes  \\
    terminal & & & & & \\
  \midrule
    matches graph & Good & Good & Poor & Good & Poor \\
    semantics & & & & & \\
  \midrule
    compact & Good & N/A & Good & Fair & Fair \\  
  \midrule
    interactive & Yes & No & No & No & Yes\\
  \midrule
  \bottomrule
  \end{tabu}
\end{table}

\section{The Spack Package Management System} 
\label{sec:background}

We provide background on package management systems and in particular,
Spack~\cite{Gamblin2015Spack}, the package management system used in this
work. We describe the current visualization support in Spack. Finally, we
perform a task analysis on package dependency graphs that leads to our
visualization design.

Package management systems (or {\em package managers}) are software tools that
aim to ease software installation and maintenance. The term {\em package}
refers to a particular software application, its related digital artifacts,
and the information necessary to automatically install, update, and configure
it\footnote{The term {\em package} is also used in languages such as Java to
describe class organization. We do not address that type of package in this
work.}. Such information includes the other software which must already be
installed. We refer to these required software packages as the target package's {\em
dependencies}.  When installing a package, the package management system will
traverse the dependency information and install any dependencies required to
build and run the target software.

A package's dependencies may themselves have dependencies, 
\begin{wrapfigure}{r}{1.8cm}
\centering
\vspace{-7pt}
\includegraphics[width=0.15\columnwidth]{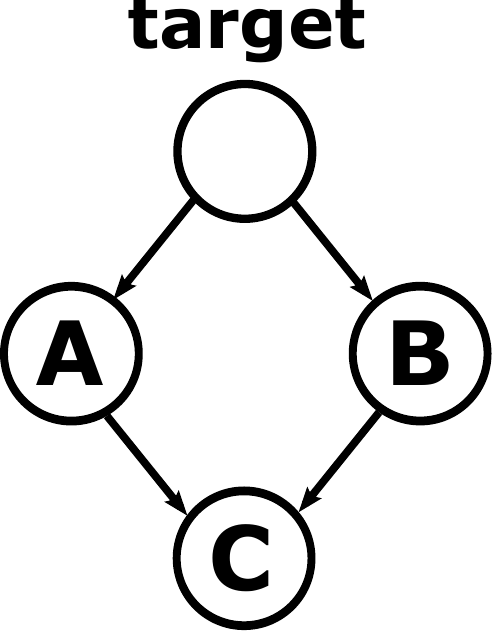}
\vspace{-7pt}
\caption{Dependency graph of target.}
\label{fig:basicdag}\vspace{-14pt}
\end{wrapfigure}
some of which may
overlap with each other. Thus, the relations 
among a package and its
dependencies form a directed acyclic graph (DAG). Fig.~\ref{fig:basicdag} shows
a small example in which the target package directly depends on packages A and
B. Packages A and B in turn both depend on package C. This is known as a
{\em diamond dependency}. In the simple case, the package manager need only to
ensure packages A, B, and C are installed before installing the target
package. However, complications arise when packages A and B require different
versions of package C. 

Spack is a package management system designed to streamline the process of
building software when multiple versions of dependencies may be needed. It is
motivated by scientific software on supercomputers---shared systems where
different users have different requirements not only in terms of software
versions, but also in the compilers and build options used.  Spack makes it
easier to concurrently build and maintain packages that depend on different
versions and build configurations of the same software.  

Dependencies in Spack may be fulfilled by one of multiple other packages. For
example, many supercomputing programs depend on the message passing interface
(MPI)~\cite{mpi31}. Several implementations of this interface exist. Spack may
choose one or the user may specify which one Spack should build against.
System-dependent options, such as MPI availability, along with updates to the
package database require users to regenerate dependency graphs they wish to
analyze to ensure up-to-date versions that reflects the state of their system.

Access to supercomputing resources is typically via shell, a command-line
interface. Developing and testing on supercomputers is essential as
supercomputers are the primary targets of large-scale scientific software.
Thus, package creation, maintenance, debugging, and distribution is done
entirely at the command line, as described by the Spack package creation
tutorial~\footnote{http://spack.readthedocs.io/en/latest/packaging\_guide.html}.

\subsection{Visualizing Dependencies in Spack}
\label{sec:git}

\begin{figure}[t]
    \centering
    \includegraphics[width=1.0\columnwidth]{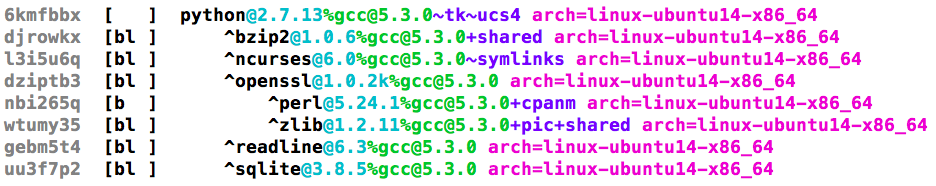}
    \caption{ASCII indented tree showing python dependencies in Spack. Each
    line includes the hash of the package configuration, the dependency type,
    and the package name, 
    version, compiler, configuration, and architecture.
    }
    \label{fig:pythontree}
    \vspace{-0.05in}
\end{figure}

The Spack \texttt{spec} command shows dependencies of a package as an indented
ASCII tree. When more information about the build, such as versions and
compilers, are known, these are shown in the tree as well
(Fig.~\ref{fig:pythontree}). The tree view emphasizes what a package {\em
depends on} (i.e., the package's dependencies). In many situations this is
satisfactory.  However, some tasks (described in the next section) benefit
from viewing the other direction---which packages are {\em affected by} a
package (i.e. those packages which depend on the target packages). Sometimes
an even a more general overview of the dependency structure is desired.  In
these cases, Spack augments the dependency tree by providing two utilities for
inspecting dependency graphs: a \texttt{dot} format description of the graph
and an ASCII representation based on the \texttt{git log --graph} command. We
briefly describe the \texttt{git log --graph} command and the pre-existing
Spack ASCII representation.

\begin{figure}[t]

\subfloat[\label{fig:git:log}]{
    \includegraphics[width=0.6\columnwidth]{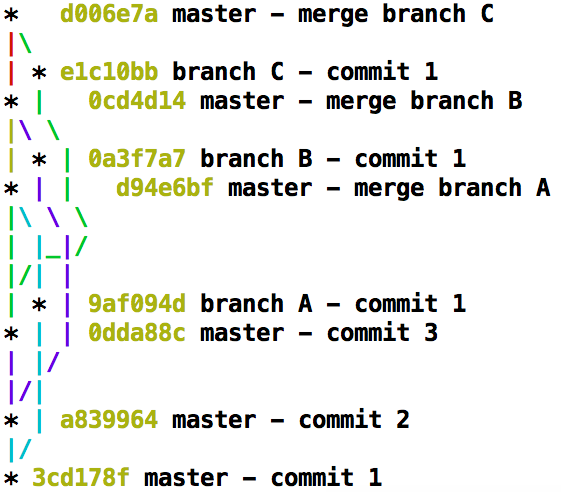}
}
\hspace{1ex}
\subfloat[\label{fig:git:python}]{
    \includegraphics[width=0.25\columnwidth]{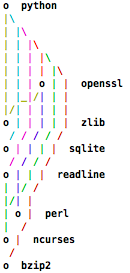}
}
    \caption{(a) Result of \texttt{git log --graph --online --all} and (b)
\texttt{git}-style python dependency graph.}
    \label{fig:gitlog}
    \vspace{-0.05in}
\end{figure}

The \texttt{git} command \texttt{git log --graph} provides an ASCII
representation of repository commits and their branching behavior. It places no more
than one commit per row with the most recent commit (of the current branch) at
the top. The sequential connections between two commits are shown using
edges drawn using \texttt{|}, \texttt{\_}, \texttt{/}, and
\texttt{$\backslash$}. Fig.~\ref{fig:git:log} shows an example. The left-most
vertical line shows commits to the master (initial) branch. Three more
branches are created and merged.

Spack's graph command adapts the
\texttt{git log --graph} algorithm~\cite{gitloggraph} to show the package
dependency graph. Unlike commits, package dependencies do not have a
temporal order, so a topological sort is used instead.
This places the package of interest on the first row with
its direct dependencies on the following rows. Fig.~\ref{fig:git:python} shows
the Spack dependency graph of python drawn with this \texttt{git}-like
approach.

The edge colors denote which dependency the edge leads to. This encoding helps
users track an edge across several rows. The sixteen ANSI colors (the eight
original and eight high intensity versions) are assigned to the packages in a
round-robin fashion. 

\begin{figure}[t]
    \centering
    \includegraphics[width=1.0\columnwidth]{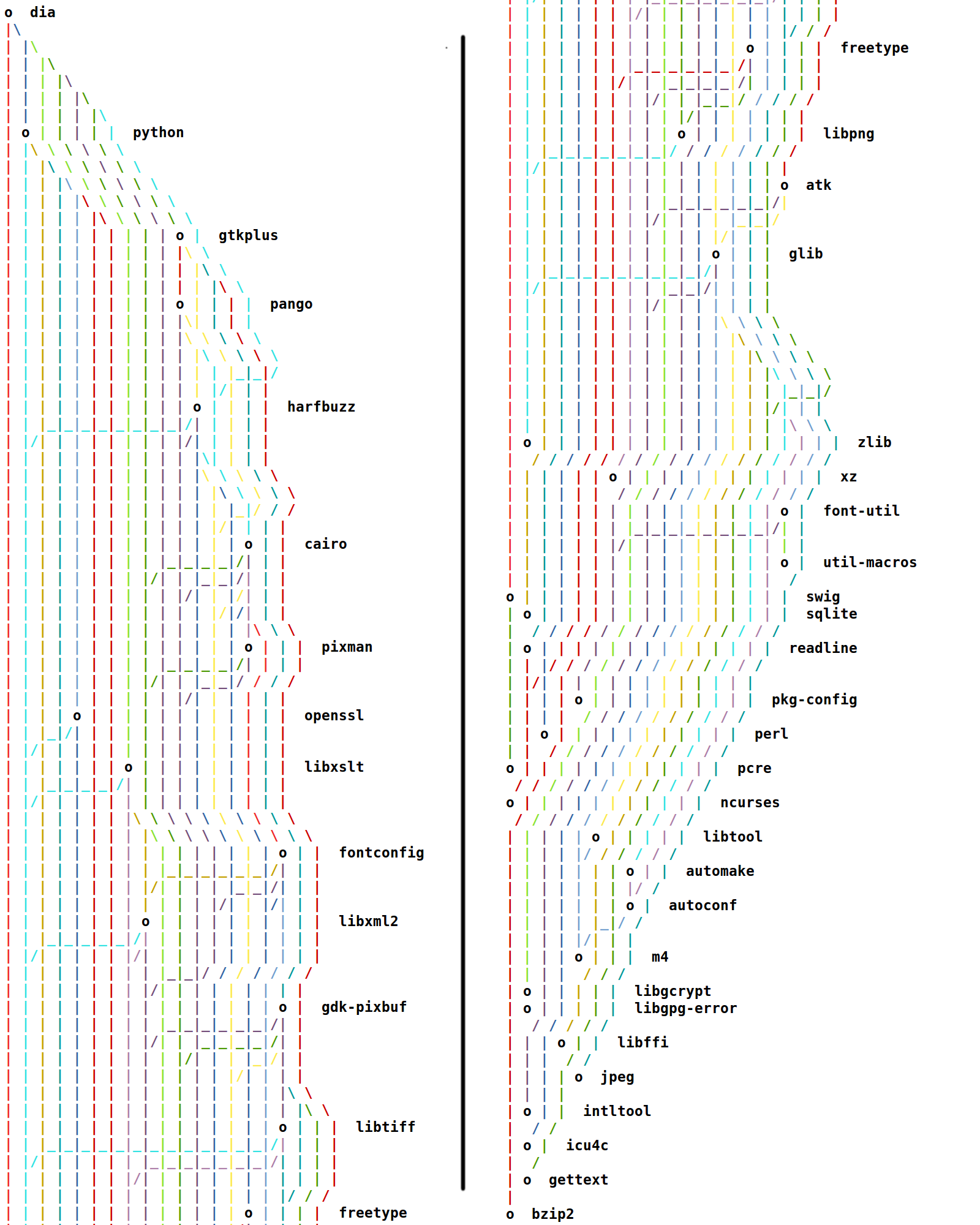}
    \caption{\texttt{git}-style package dependency graph of \texttt{dia} (also
    shown in Fig.~\ref{fig:teaser}). The
    \texttt{freetype} node has been duplicated to show alignment between the
  two halves.}
    \label{fig:git-dia}
    \vspace{-0.05in}
\end{figure}

There are two major advantages to this \texttt{git}-like layout. First, the
layout conserves columns, thus fitting in the 80 column limits preferred by
some command line users.  The conservative use of columns can also be a
downside as the resulting graph is visually dense.  Second, the
\texttt{git}-like layout is an unambiguous representation: when one edge is
routed into another, they both exit at the same terminus.

While this heavily vertical style matches well with the pure text \texttt{git
log} command, the temporal nature of \texttt{git} commits, and the propensity
for long chains in the resultant graph, it obscures the layered nature of
dependency graphs. Furthermore, the \texttt{git}-like graphs require many
rows, requiring users to scroll even for relatively small graphs. The Spack
maintainers we collaborated with (see Sec.~\ref{sec:taskabstraction}) wanted a
more compact representation that used fewer rows.

Fig.~\ref{fig:git-dia} shows the \texttt{git}-like dependency graph for
\texttt{dia}, a package with 39 dependencies (also depicted using our approach
in Fig~.\ref{fig:teaser}).  Visually tracking some edges can require several
page-up operations.  While the edge coloring can help users keep their place,
as these are assigned before edge layout, sometimes the same or similar colors
cross or appear side-by-side as with the two long red edges in
Fig.~\ref{fig:git-dia}.

\subsection{Task Analysis and Abstraction}
\label{sec:taskabstraction}

We seek to improve the dependency graph visualization for Spack. First, we
consider how dependency graph visualizations are used. Through a series of
interviews with two Spack maintainers (one via a video-conference, four text
chats, and an informal in-person discussion, the other via an informal
in-person discussion), we identified three user groups and their tasks with
respect to dependency graphs. We focus specifically on graph-related tasks, as other
tasks, such as simply viewing the set of package dependencies or viewing only
what a particular package depends on (rather than what depends on it),  are
already well supported by Spack's indented tree listing
(Fig.~\ref{fig:pythontree}).

\vspace{0.5ex}

\noindent\textbf{Audience and Their Tasks.}
There are three classes of people who consult Spack dependency graphs: Spack
power users, package developers, and Spack maintainers/contributors.  We
discuss their goals below and relate them to the task taxonomy for graphs of Lee et
al.~\cite{Lee2006}. We summarize our classification in
Sec.~\ref{tab:taskanalysis}. This analysis was reviewed after development by
the Spack maintainer we had the most contact with, who is among the authors of
this paper.

\begin{table}
  \caption{Tasks Abstraction for Spack Dependency Graphs}
  \label{tab:taskanalysis}
  \scriptsize
  \centering
  \begin{tabu}{
      *{3}{l}
    }
  \toprule
    Graph Task & Spack Task & Spack Role \\
  \midrule
    topology -- & $\cdot$ Determine packages {\em affected by}
    & $\cdot$ users, \\
    accessibility & \hspace{0.8ex} a package & \hspace{0.8ex} developers \\

                              & $\cdot$ Identify diamond dependency & $\cdot$ maintainers \\
  \midrule
    browsing -- & $\cdot$ Find source(s) of a dependency & $\cdot$ developers \\ 
    follow path & & \\
  \midrule
    overview & $\cdot$ Assess trade-offs among options & $\cdot$ users \\
             & $\cdot$ Assess complexity to judge & $\cdot$ maintainers \\
	     & \hspace{0.8ex} performance & \\
  \midrule
    attribute-based & $\cdot$ Identify dependency type or install &
    $\cdot$ all \\
    (all) & \hspace{0.8ex} configuration & \\
  \bottomrule
  \end{tabu}
\end{table}

Spack users may refer to a package dependency graph to understand
relationships between the other packages their target software depends on and
thus what optional constraints they may want to specify, such as a particular
parallel runtime library. Knowing which other packages may be affected can be useful in
their decision-making process. Consider a power user `Yulia' trying to
install a scientific package with which she has a passing familiarity. Yulia
favors the parallel runtime implementation provided by GroupX because it has yielded
performance improvements for her on a previous project. However, she
understands other packages are known to perform better with the application
suite of GroupY. She wants to examine the graph to assess the potential
trade-offs in choosing particular implementations.  Recognizing these
situations requires identifying direct and indirect connections and gaining a
sense of how all the connections work together, in other words, ``topology --
accessibility'' tasks, ``overview'' tasks, and ``attribute-based'' tasks in
Lee et al.'s taxonomy.

In addition to performing the power user's tasks for debugging, package
developers may examine dependency graphs to verify they have included all the
necessary build information in their package. Unexpected dependencies in the
graph may indicate an inadequate specification. Tracing a path from the
dependency up to its sources, as motivated by a particular error message, can
help resolve an error. Consider a package developer `Devon' who is testing
his package on a supercomputer accessible by many of his target users. He
tries multiple possible configurations and is surprised by some of the ways in
which Spack resolves the dependencies. Devon consults the dependency graph to
understand why some configuration choices affect other packages in the graph.
These tasks utilize both the identification of direct and indirect connections
(task: topology -- accessibility) as well as following a path (task: browsing
-- follow path).

Spack maintainers analyze package dependency graphs when debugging, adding
features, or testing. Consider a Spack maintainer `Mabel' who is investigating
a report of the Spack system failing to build a package due to choices it made
while resolving dependencies. Mabel wants to check that the dependencies truly
exist and if they conflict with each other in a way Spack was unable to
determine. She consults the dependency graph, looking in particular for
diamond dependencies and other instances of multiple dependencies as these
present difficulties to the dependency resolution algorithm. This is again a
task about identifying connections (task: topology -- accessibility). 

In addition to investigating bugs, Mabel wants to evaluate the performance of
the package management system. Should she notice a particular package takes a
long time, she can look at the dependency graph to gain a sense of the
complexity of any package installation (task: overview). Information about the
version, compiler, and options selected for each package, as well as the type
of dependency (e.g., required for build only, called by the target, or linked
by the target) is also of interest (tasks: attribute-based (all)).

As all users perform topology-based tasks on dependency graphs, node-link
diagrams are an appropriate choice to represent them. We note that indented
trees may be more suitable for some Spack user scenarios where all packages
depending on a particular choice need not be found and indirect connections
need not be evaluated. These scenarios are already served by Spack's indented
tree feature, which also includes rich attribute information. The node-link
diagram augments this functionality allowing the exploration of more
complicated topology-based tasks.

\vspace{0.5ex}

\begin{figure}[h]
    \centering
    \includegraphics[width=1.0\columnwidth]{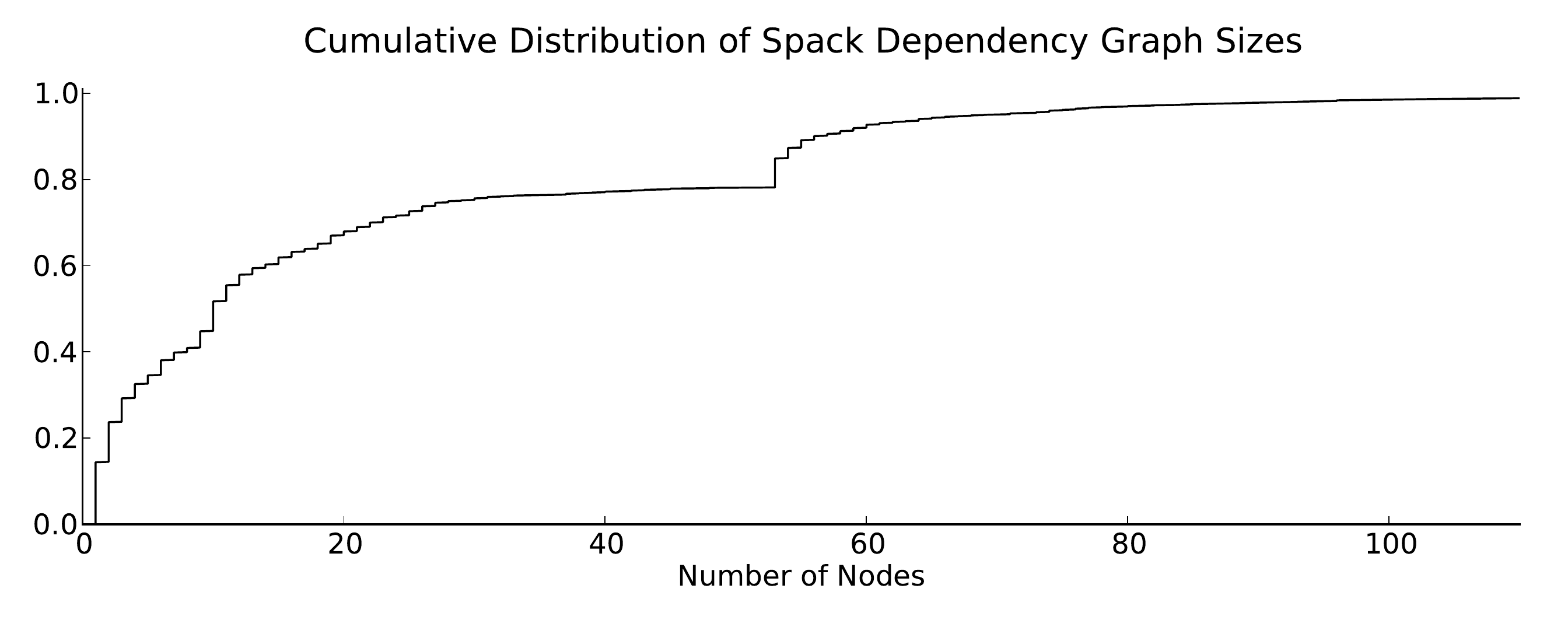}
    \caption{Most Spack dependency graphs have fewer than 50 nodes.}
    \label{fig:graphsizes}
    \vspace{-0.05in}
\end{figure}

\noindent\textbf{Data.}
There are over 2,100 packages in the Spack library as of June 2018. The
majority of them have dependency graphs with fewer than 50 nodes.
Fig.~\ref{fig:graphsizes} shows the cumulative distribution of graph sizes by
node count. Node-link diagrams are effective for representing graphs that are
relatively small in size and emphasize topology-based tasks~\cite{Ghoniem2004,
Keller2006} and thus we use them to represent package dependency graphs.

\vspace{0.5ex}

\noindent\textbf{Workflow.} 
Spack is a command line tool. Spack users install packages through the command
line. Package developers use the Spack commands \texttt{spack create} and
\texttt{spack edit} at the terminal to create and maintain their packages and
then test them on their target supercomputers via remote login.  Similarly
Spack maintainers test and debug the system on the command line via the same
interface. As one of the motivations for Spack was streamlining the build process
on supercomputing systems, much of this command line access is to a remote,
and often secure, system that may not have graphical applications such as a
graphical web browser installed. Furthermore, users may have limited
privileges and be restricted to command line access. 

Typically, to view a graphical visualization, users must shift focus back to
their local machine, copy the file from the supercomputer, and launch a local
viewer. This adds several steps to the process and takes the user away the
rest of their analysis. Utilities like \texttt{rsync} can streamline the file
copy process when many packages need to be analyzed in a workflow session.
While this may be the case for Spack maintainers, it is not so for users and
package developers. In the case that graphical applications are available, the
user may view them via X11 forwarding {\em if} they had the foresight to login
with that option enabled. Depending on the system and their location, this
option can induce significant lag. For example, we launched a PDF viewer with
a graph through X11 forwarding on one of our users' systems and incurred a
penalty of greater than 10 seconds to launch the viewer and then on the order
of 1 second for operations such as redrawing for panning, expanding a
drop-down menu, and re-sizing.

During our interviews, the users expressed the desire for an ASCII-based
visualization for use on the command line that was more
`\texttt{dot}'-like than their current one~(\ref{sec:git}). We discussed the
possibilities of an interactive, browser-renderable tool, especially to support
the multi-variate attribute-based tasks, but reception to that proposal was
lukewarm. The users were more interested in a console-based tool that would
help the majority of their smaller analysis tasks first. Given the command
line workflow, the overhead of one-off copying of files, the barriers to using
graphical tools, the small size of the graphs, and the need to support
topological operations, we agreed that initial request of the users was viable
and proceeded to design an improved visualization of package dependencies as
ASCII node-link diagrams.

\section{\texttt{graphterm}}
\label{sec:ascii}

To support the command-line workflow for Spack community members, we developed
\texttt{graphterm}~\footnote{http://github.com/kisaacs/graphterm}, a
Python tool for interactive ASCII dependency graph visualization.
\texttt{graphterm} adapts a layered graph layout that uses graphical marks to
create a layout using ASCII characters. 
Unlike the \texttt{git}-like layout, \texttt{graphterm} bundles edges,
resulting in ambiguity. Users can resolve the ambiguity via interactive
highlighting. We describe the design and implementation of the layout of
\texttt{graphterm} in Sec.~\ref{sec:graphterm:layout}. We then discuss the
supported interactivity in Sec.~\ref{sec:graphterm:interactions}. 

\begin{figure*}[t]
    \centering
    \subfloat[\texttt{mkfontdir} graph with graphical marks.\label{fig:mkfontdir:tulip}]{
    \includegraphics[width=1.2\columnwidth]{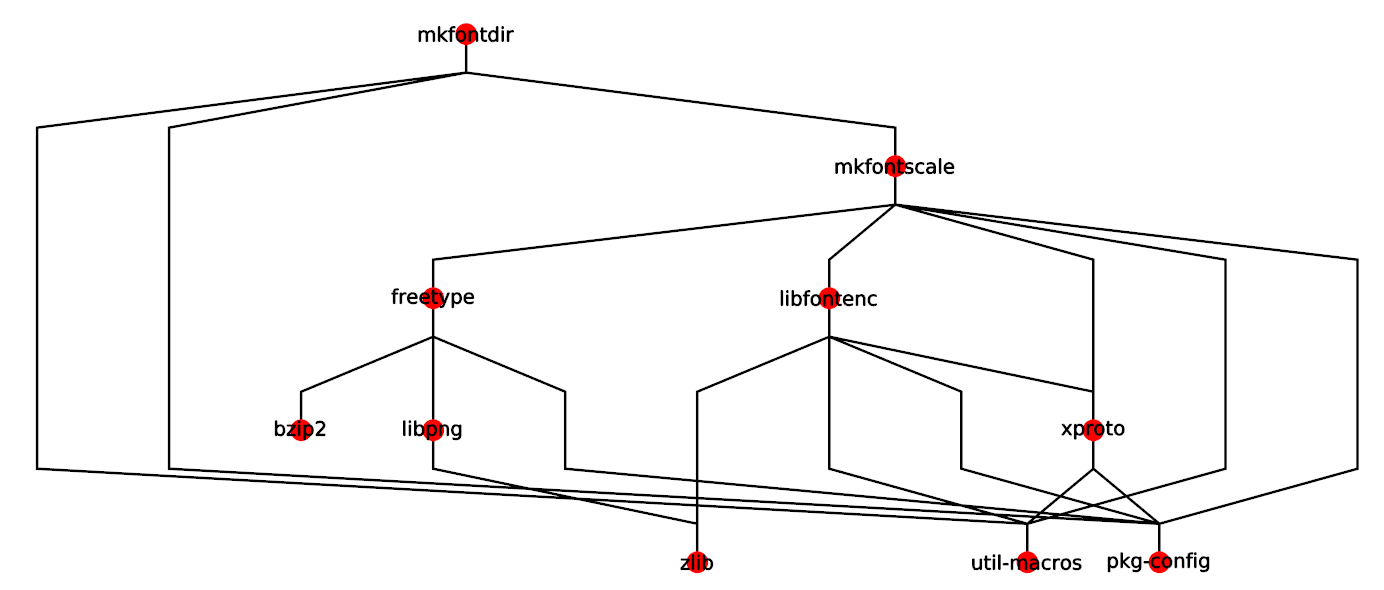}
}
\hspace{2ex}
\subfloat[\texttt{mkfontdir} graph with ASCII marks.\label{fig:mkfontdir:term}]{
    \includegraphics[width=0.74\columnwidth]{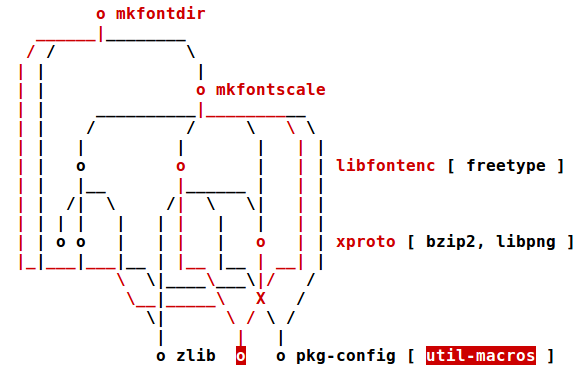}
}
\caption{The incoming edges of \texttt{util-macros} and \texttt{pkg-config}
cross several others in the original layout. We re-route those that cross the
vertical edge of \texttt{zlib} to be horizontal but vertically offset from
each other by package. \texttt{util-macros} and its neighbors are highlighted
  in the \texttt{graphterm} depiction, demonstrating the depiction of the incoming edge from
  \texttt{mkfontdir}.}
    \vspace{-0.05in}
\label{fig:mkfontdir}
\end{figure*}

\subsection{\texttt{graphterm} Layout}
\label{sec:graphterm:layout}

From our discussions with Spack maintainers (Sec.~\ref{sec:taskabstraction}) we
concluded that the users conceptualize dependency graphs in a fashion similar
to a layered graph layout~\cite{Sugiyama}.  One Spack maintainer specifically
requested a ``\texttt{dot}-like'' layout in ASCII. Furthermore, we concluded
that such a layout supports their tasks. Users can quickly determine the
direction of dependencies by vertical order, unlike in an orthogonal layout
which may be more adaptable to ASCII, but does not maintain a vertical or
horizontal ordering of nodes.

The layout algorithm (Algorithm~\ref{alg:graphterm}) starts by obtaining a
graphical layered layout. Based on the mark placement and induced crossings
therein, it generates a corresponding set of node and edge positions on a grid
(Algorithm~\ref{alg:bundleedges}, Algorithm~\ref{alg:togridpoints}). Finally,
ASCII characters from the set $\{$ \texttt{|}, \texttt{\_}, \texttt{/},
\texttt{$\backslash$}, \texttt{o}, \texttt{X} $\}$ are placed to represent
those nodes and edges (Algorithm~\ref{alg:placemarks}). We then place node
labels to fit (Algorithm~\ref{alg:placelabels}).

\begin{algorithm}
    \small
    \SetAlgoLined\Indmm
    \SetKwFunction{GraphTerm}{graphterm} 
    \SetKwFunction{TulipHierarchical}{Tulip\_hierarchical} 
    \SetKwFunction{GetCrossings}{get\_edge\_crossings} 
    \SetKwFunction{SetInsert}{insert} 
    \SetKwFunction{BundleEdges}{get\_bundle\_positions} 
    \SetKwFunction{ToGridPoints}{to\_grid\_points} 
    \SetKwFunction{PlaceOnGrid}{place\_on\_grid} 
    \SetKwFunction{PlaceLabels}{place\_labels} 
    \SetKwBlock{Begin}{\vspace{-3mm}}{}%
    \GraphTerm(G)\;
    \Begin{
      positions = \TulipHierarchical{\em G}\;
      crossings = \GetCrossings{\em positions}\;
      positions += \BundleEdges{\em G, crossings}\tcp*{Algorithm 2}
      xset = [], yset = []\;
      \lFor{{\em x, y} in {\em positions}} {
	xset.\SetInsert{\em x},
	yset.\SetInsert{\em y}
      }
      rows, columns = \ToGridPoints{\em G, xset, yset}\tcp*{Algorithm 3}
      grid = \PlaceOnGrid{\em G, columns, rows}\tcp*{Algorithm 4}
      grid = \PlaceLabels{\em G, grid}\tcp*{Algorithm 5}
    }
    \caption{GraphTerm Layout Overview}
    \label{alg:graphterm}
\end{algorithm}

\vspace{1ex}

\noindent\textbf{Converting graphic layout to an ASCII grid.}
We considered several existing hierarchical and layered graph layouts to adapt
to ASCII. The considered layouts included \texttt{dot},
\texttt{dagre}~\cite{dagre}, and the hierarchical layouts included in
Tulip~\cite{Auber2004} and OGDF~\cite{OGDF}.  We chose the Tulip hierarchical
layout~\cite{Auber2004} as a basis because it is a straight-edge layout that
uses mostly vertical and diagonal edges with a propensity to re-use vertical
edges in our package dependency graphs. This resulted in less clutter in
comparison to other layouts as well as marks that were easier to adapt to
ASCII. 

We run the Tulip hierarchical layout\footnote{The figures and study in this
paper use the Tulip Python bindings. The version released on Github ports the
layout into pure Python. The port differs in some of its sort orders and
vertical spacings and thus can produce a slightly different ASCII layout.} on
the package dependency graph. From the graphical layout, we obtain initial
positions for all nodes as well as line segment end points for all edges.

\begin{figure}[t]
    \centering
    \includegraphics[width=1.0\columnwidth]{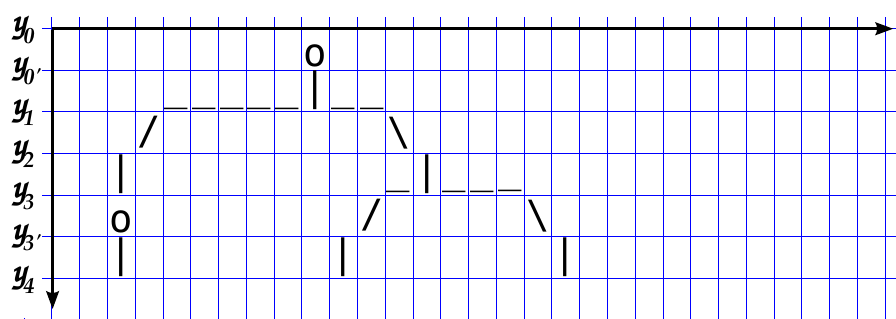}
    \caption{We assign values from our position set to the upper left corner
    of each character cell. Rows with nodes are assigned an extra $y$ value.
    In this fictional example, the top node spans grid lines $y_0$ and $y_{0'}$ and the
    second node spans $y_3$ and $y_{3'}$. No other rows are doubled.}
    \label{fig:gridnumbering}
    \vspace{-0.05in}
\end{figure}

First, we determine sets of $x$ and $y$ positions of note in the layout---the
positions of the nodes, the end points of the line segments, and the positions
of the crossings. The layered nature of the layout induces a small set of $y$
values.  In the Tulip hierarchical layout, dummy nodes are inserted into the
graph at the pre-existing layers to aide in edge routing and within-layer node
ordering. The location of the dummy nodes correspond to the non-node endpoints
of the line segments that compose each edge. Combined with the true nodes,
these induce a small set of $x$ values. We will ultimately convert these
floating point $x$ and $y$ values to a discrete compact grid
(Fig.~\ref{fig:gridnumbering}), but first we must add values to the sets to
account for edge crossings.

We compute the location of all line segment crossings in the graphical layout.
In an early design, we split all segments at their crossings and added the
crossing $x$ and $y$ positions to our sets. While this served to spread out
dense sections of the graph, thereby making it possible to follow each edge,
it also expanded the needed grid size unnecessarily. The afforded space for
dense crossings also detracted from communicating the overall structure of the
graph. Thus, in regions with a large number of crossings, we selectively
re-route segments to share crossing points. We found a good heuristic was to
re-route the incoming diagonal edges of (dummy) nodes.

\vspace{1ex}

\noindent\textbf{Adjustments for edge re-use and bundling.}
Package dependency graphs often have a few nodes with either a high in-degree
or a high out-degree. In our chosen layout, this results in several diagonal
segments fanning in or out between two layers. Many of the edge crossings are
caused by these structures.  The large number of segments with slightly
different angles is also difficult to represent with our limited set of glyphs
(ASCII). In the absence of edge crossings, our choice of line drawing heuristic, 
described later in this section, results in
re-use of horizontal segments created by underscores, similar to the re-use of
vertical segments in the original graphical layout.

When we detect crossings between a diagonal segment and a vertical segment,
rather than adding the crossing position to our $x$ and $y$ sets, we alter the
$y$ crossing position to a set value based on the end (dummy) node of the
segment. The altered $y$ position, calculated in lines 7-8 of
Algorithm~\ref{alg:bundleedges}, uses the end node's $x$ position to guarantee
uniqueness in the presence of other such nodes with the same $y$ value,
choosing a position (\texttt{routed\_y}) between the node and the top of the
segment (\texttt{segment.y1}). The procedure routes all diagonal segments to
that (dummy) node through the same $y$ value, thus avoiding increasing our set
of $y$ values for each crossing. In the case of crossings between two diagonal
segments, if such a $y$ value has been set by a diagonal-vertical crossing, we
use it. If two such $y$ values exist, we omit the crossing. Otherwise, we use
the computed (true) crossing value. 

\begin{algorithm}[t]
    \small
    \SetAlgoLined\Indmm
    \SetKwFunction{BundleEdges}{get\_bundle\_positions}
    \SetKwFunction{Endpoints}{get\_endpoints}
    \SetKwFunction{Map}{map}
    \SetKwFunction{EmptyMap}{empty\_map}
    \SetKwFunction{InSegments}{in\_segments}
    \SetKwFunction{CrossesVertical}{crosses\_vertical}
    \SetKwFunction{ReRoute}{shift\_crossings}
    \SetKwFunction{Break}{break}
    \SetKwBlock{Begin}{\vspace{-3mm}}{}%
    \BundleEdges(G, crossings)\;
    \Begin{
      dummy\_nodes = \Endpoints{\em G.links.segments}\;
      all\_nodes = \{G.nodes, dummy\_nodes\}\;
      bundle\_points = \EmptyMap{}\;
      \For{{\em node} in {\em all\_nodes}} {
	\For{{\em segment} in \InSegments{\em node}} {
	  \If{\CrossesVertical{\em segment}} {
	    offset\_factor = 0.5 $\times$ node.x $/$ G.max\_x\;
	    routed\_y = (node.y - segment.y1) $\times$ offset\_factor\;
	    \ReRoute{\em \InSegments{\em node}, routed\_y, crossings}\;
	    \Break{}\;
	  }
	}
      }
      \Return{crossings}
    }
    \caption{GraphTerm Edge Bundling: Shifts crossings of a node's incoming
    segments if any cross another vertical segment.}
    \label{alg:bundleedges}
\end{algorithm}

Fig.~\ref{fig:mkfontdir} shows our re-routing rules as applied to the
dependencies of \texttt{mkfontdir}. Several edges from the left cross the
vertical edge into \texttt{zlib}. Our algorithm re-routes them, resulting in
horizontal edges at different heights to \texttt{util-macros} and
\texttt{pkg-config}.  However, the diagonal edges crossing directly below
\texttt{xproto} are not re-routed. We found applying the re-routing to those
edges resulted in dense and overly boxy graphs that resembled grids. Our
policy was chosen to balance the compactness of the depiction with
readability.

\vspace{1ex}

\noindent\textbf{Gridding and Character Set.}
Having calculated the set of unique $x$ and $y$ values representing (dummy)
nodes and crossings, we assign each value to a row or column of a grid. This grid will
become our ASCII representation. Conceptually, we consider the upper left
corner of each monospaced character cell to be a grid coordinate. We then
chose ASCII characters to represent edges between them, deciding on the set
$\{$ \texttt{|}, \texttt{\_}, \texttt{/}, \texttt{$\backslash$}, \texttt{X}
$\}$, preferring the underscore to the dash because its end point is closer to the
grid corner. As the crossings are at the grid corners, we do not use
\texttt{+} which would be a crossing mid-cell.

A mark cannot be drawn at the corner of four character spaces, so we place the
mark for a node (\texttt{o}) in the cell itself. Therefore, any $y$ value
associated with at least one node is given two grid spaces.
Fig.~\ref{fig:gridnumbering} demonstrates this assignment.

\begin{algorithm}[t]
    \small
    \SetAlgoLined\Indmm
    \SetKwFunction{ToGridPoints}{to\_grid\_points}
    \SetKwFunction{Sort}{sort}
    \SetKwFunction{Map}{map}
    \SetKwFunction{EmptyMap}{empty\_map}
    \ToGridPoints(G, xset, yset)\;
    \SetKwBlock{Begin}{\vspace{-3mm}}{}%
    \Begin{
      xlist = \Sort{\em xset}, ylist = \Sort{\em yset}\;
      columns = \EmptyMap{}, rows = \EmptyMap{}\;
      col = 0, row =0\;
      \lFor{{\em x} in {\em xlist}} {
	columns.\Map{\em x, col},
	col += 2;
      }
      \For{{\em y} in {\em ylist}} {
	rows.\Map{\em y, row}, row += 2\;
	\lIf{$\exists $ {\em node} in {\em G} $\mid$ {\em node.y = y}} {
          row += 2
	}
      }
      \Return{rows, columns}
    }
    \caption{Translate Real Positions to Grid Points}
    \label{alg:togridpoints}
\end{algorithm}

To keep the ASCII layout compact, rather than maintain the relative distances
between positions, we assign the values in order in our grid
coordinates subject to some expansion function. In our implementation, we use
a $2\times$ multiplier to prevent the graph from becoming too dense, as shown
in Algorithm~\ref{alg:togridpoints}. Therefore each $x$ value is assigned to
successively numbered even columns. The $y$ values are assigned similarly with
the added row for nodes described above.

\vspace{1ex}

\noindent\textbf{Edge Layout.}
Once the correspondence between graphical layout positions and grid points are
set, we assign ASCII characters to the grid cells. Note that segments may span
many grid points and thus many grid cells. We assign nodes (\texttt{o}) and
purely vertical segments first. Vertical segments require only successive
vertical bar (\texttt{|}) characters. 

\begin{figure}[t]
    \centering
    \includegraphics[width=0.98\columnwidth]{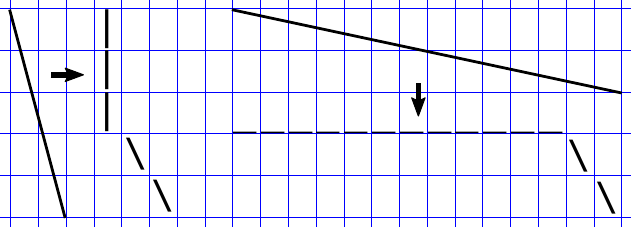}
    \caption{Graphical lines are converted to ASCII first by excess
    vertical or horizontal displacement then by diagonal displacement. 
    The left line is more vertical and thus has no underscores. The right line is
    more horizontal and thus has no vertical bars.}
    \label{fig:asciilines}
    \vspace{-0.05in}
\end{figure}

There are several options for drawing the non-vertical segments. We initially
used the grid cells calculated by Bresenham's line drawing
algorithm~\cite{Bresenham1965}, but this results in unnecessarily crooked
``lines'', clutter, and cell collisions.  Instead, we break each segment into
$0^{\circ}$ (horizontal), $45^{\circ}$/$135^{\circ}$ (diagonal), and
$90^{\circ}$ (vertical) pieces summing to the effective displacement. Edges
that traverse more horizontally have horizontal and diagonal pieces but no
vertical pieces. Edges that traverse more vertically have diagonal and
vertical pieces but no horizontal pieces. We draw either the excess horizontal
(with underscores in the cell above) or vertical (with vertical bars) first,
then the diagonal with slashes. Examples are shown in
Fig.~\ref{fig:asciilines}. The translation is described in
Algorithm~\ref{alg:placemarks}.  This drawing scheme leads to straighter
segments, which have been shown to support path finding~\cite{Ware2002}. It
also tends to naturally overlap edges, effectively coalescing edge marks along
main horizontal and vertical thoroughfares.

\begin{algorithm}[t]
    \small
    \SetAlgoLined\Indmm
    \SetKwFunction{PlaceOnGrid}{place\_on\_grid}
    \SetKwFunction{Initialize}{initialize\_to\_spaces}
    \SetKwFunction{Place}{place}
    \SetKwFunction{TransformCoords}{transform\_coords}
    \SetKwFunction{Abs}{abs}
    \SetKwFunction{Max}{max}
    \SetKwFunction{Min}{min}
    \SetKwFunction{Sign}{sign}
    \SetKwBlock{Begin}{\vspace{-3mm}}{}%
    \PlaceOnGrid(G, columns, rows)\;
    \Begin{
      order = \{`$\mid$': 1, `\_': 2, `\textbackslash': 3, `/': 3, `X': 0, ` ': 4\}\;
      grid = \Initialize{}\;
      \lFor{{\em node} in {\em G}} {
	grid.\Place{'o', {\em columns[node.x]}, {\em rows[node.y]}}
      }
      \For{{\em segment} in {\em G.links}} {
	r, c, r2, c2 = \TransformCoords{\em segment, columns, rows}\;
	diagonal = \Min{\em (r2 - r), \Abs{\em c2 - c}}\;
	vertical = \Max{0, {\em (r2 - r) - \Abs{\em c2 - c}}}\;
	\lIf{\em c $<$ c2} {
	  slash = `\textbackslash';
	  cdiag = c2 + diagonal
	}
	\lElse{
	  slash = `/';
	  cdiag = c2 - diagonal
	}
	\eIf{\em vertical $>$ 0} {
	  \For{{\em i = r} to {\em r + vertical}} {
	    \lIf{\em order[`$\mid$'] $<$ order[grid[c][i]]} {
	      grid.\Place{`$\mid$', {\em c, i}}
	    }
	  }
	  r = r + vertical + 1\;
	}
	{
	  \For{{\em j = c} to {\em cdiag}} {
	    \lIf{\em order[`\_'] $<$ order[grid[j][r - 1]]} {
	      grid.\Place{`\_', {\em j, r - 1}}
	    }
	  }
	  c = cdiag + sign(c2 - c)\;
	}
	\For{{\em i = r} to {\em r2}} {
	  \lIf{\em order[slash] $<$ order[grid[c][i]]} {
	    grid.\Place{\em slash, c, i}
	  }
	  \If{\em order[slash] = order[grid[c][i]] and grid[c][i] $\neq$ slash} {
	    grid.\Place{`X', {\em c, i}}\;
	  }
	  c += sign(c2 - c);
	}
      }
      \Return{grid}
    }
    \caption{Place ASCII Marks in Grid}
    \label{alg:placemarks}
\end{algorithm}

Assigning ASCII characters to connect the segments can result in collisions.
Different segments may require a different ASCII character in the same grid
cell. We appeal to the Gestalt principle of continuation to resolve these
collisions. Using line breaks in edges at crossings in this manner has
been previously shown to have little effect on readability~\cite{Rusu2011}.
We observed that giving precedence to slashes over vertical bars and vertical
bars over underscores works to preserve segment continuity. When two opposing
slashes conflict, we use an \texttt{X} character (as seen in
Fig.~\ref{fig:mkfontdir:term}).

\vspace{1ex}

\noindent\textbf{Labels.}
After the edges have been drawn in the grid, we place the labels.  Ideally,
the labels would be placed close to their node. However, we consider the sense
of graph structure and the compactness of its representation higher
priorities. If there is enough empty space to the left or the right of the
node, we place the package name in that space. Preference is given to the
right side to match the \texttt{git}-like depiction (Sec.~\ref{sec:git}) where
all labels are on the right. Also, this allows the user to follow a link to a
node (\texttt{o}) and then continue reading from left to right to see the
label.

If there is not enough space for the package name, we place the label to the
right of the entire graph. The right-most unlabeled node per row is drawn next
to the graph. The rest are drawn in a bracketed list to the right of that
label in the left-right order of the nodes. Brackets are used to distinguish
this list from the other labels. The rightward placement is again in deference
to the \texttt{git}-like layout. The procedure is outlined in
Algorithm~\ref{alg:placelabels}. 

We considered balancing the extra labels on the
left and right side, based on their position in the grid.  However, this moved
the structure of the graph further to the right which would require users to
shift focus from the cursor which rests in the lower left corner. 

\begin{figure}[t]
    \centering
    \includegraphics[width=0.98\columnwidth]{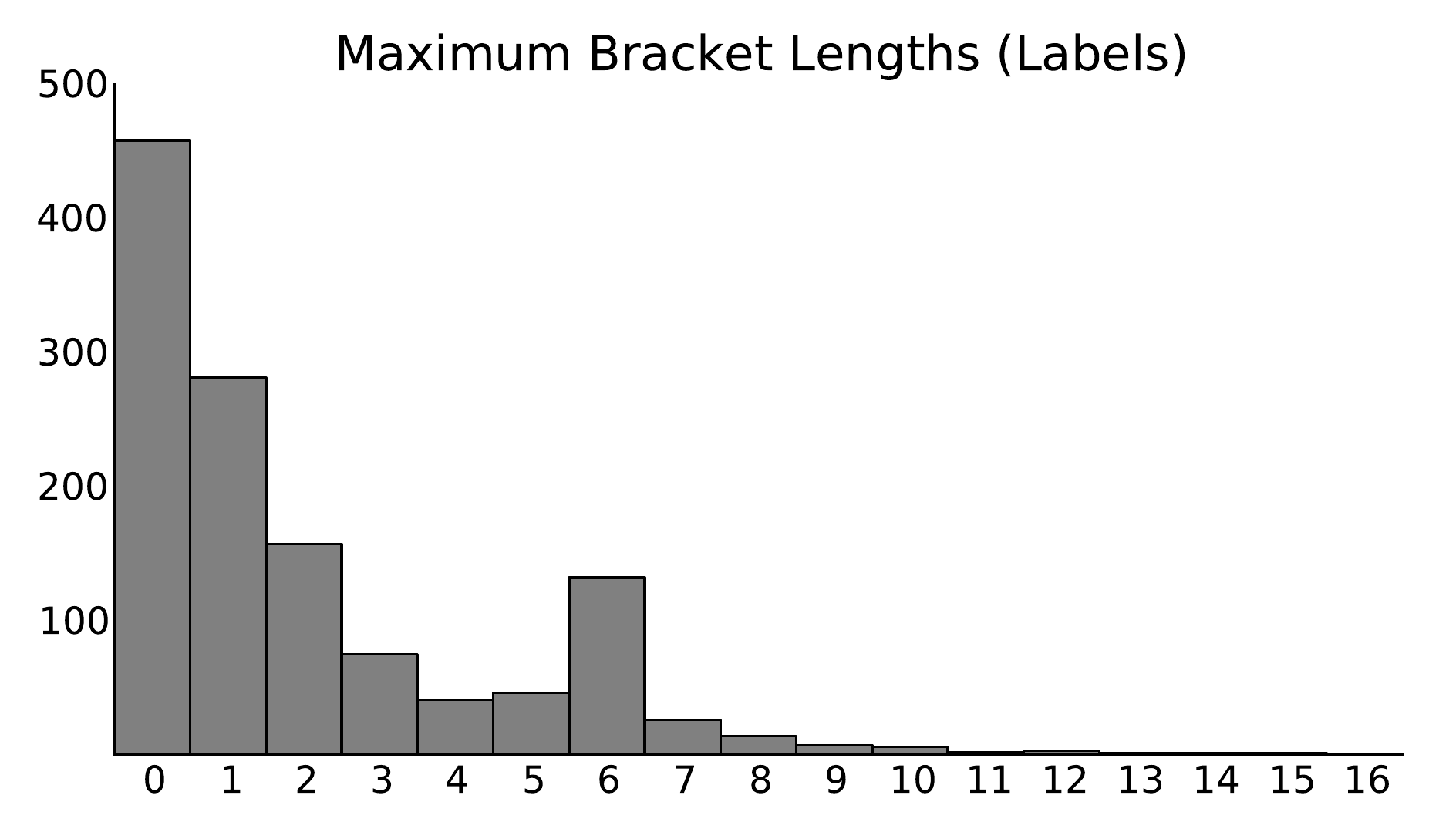}
    \caption{Histogram showing the distribution of the maximum number of
    labels relegated to the bracketed list amongst the Spack package 
    dependency graphs. There is a peak at six due to a large number of R
    libraries. At the time of this experiment, the R package, a subgraph of
    these libraries, produced a layout with a length six list.}
    \label{fig:bracketlength}
    \vspace{-0.05in}
\end{figure}

We remark the bracketed list is not ideal, but a trade-off made to emphasize
graph topology. One problem is that long lists will be truncated by the edge
of the terminal window and require panning. While one could construct a
package with an arbitrarily long list of nodes on the same level, in practice
we observed the length of the longest list grew with the number of nodes in
the graph and most graphs had a maximum list of six or fewer labels (median: 1
label, average: 2 labels).  Fig.~\ref{fig:bracketlength} is a histogram of
maximum bracket length across Spack packages.

\begin{algorithm}[t]
    \small
    \SetAlgoLined\Indmm
    \SetKwFunction{PlaceLabels}{place\_labels}
    \SetKwFunction{Continue}{continue}
    \SetKwFunction{PlaceRight}{place\_right\_of\_node}
    \SetKwFunction{PlaceLeft}{place\_left\_of\_node}
    \SetKwFunction{PlaceEnd}{append\_to\_right\_bracket}
    \SetKwFunction{LastNode}{is\_last\_in\_row}
    \SetKwFunction{PlaceRightEnd}{place\_right\_of\_graph}
    \SetKwBlock{Begin}{\vspace{-3mm}}{}%
    \PlaceLabels(G, grid)\;
    \Begin{
      \For{{\em row} in {\em grid}} {
	\For{{\em node} in {\em row}} {
	  s = node.label\;
	  \lIf{\PlaceRight{\em s}} {
	    \Continue{}
	  }
	  \lElseIf{\PlaceLeft{\em s}} {
	    \Continue{}
	  }
	  \lElseIf{\LastNode{\em node}} {
	    \PlaceRightEnd{\em s}
	  }
	  \lElse{
	    \PlaceEnd{\em s}
	  }
	}
      }
      \Return{grid}
    }
    \caption{Place Labels}
    \label{alg:placelabels}
\end{algorithm}

\subsection{\texttt{graphterm} Interactions}
\label{sec:graphterm:interactions}

We design our interactions controls to match common command line programs.  We
do this exactly when possible or by metaphor when not.  Searching and
highlighting a specific node is done by typing the forward slash
character followed by the name, as done in \texttt{less}. We expect users
may search for a node when they are considering specifying a version or
compiler and want to consider how that choice may affect packages depending on
that node. Developers and maintainers may want to search for a node to verify
its connections when debugging.

\begin{figure}[th]
    \centering
\subfloat[Direct connections highlighted.\label{fig:nco:direct}]{
    \includegraphics[width=1.0\columnwidth]{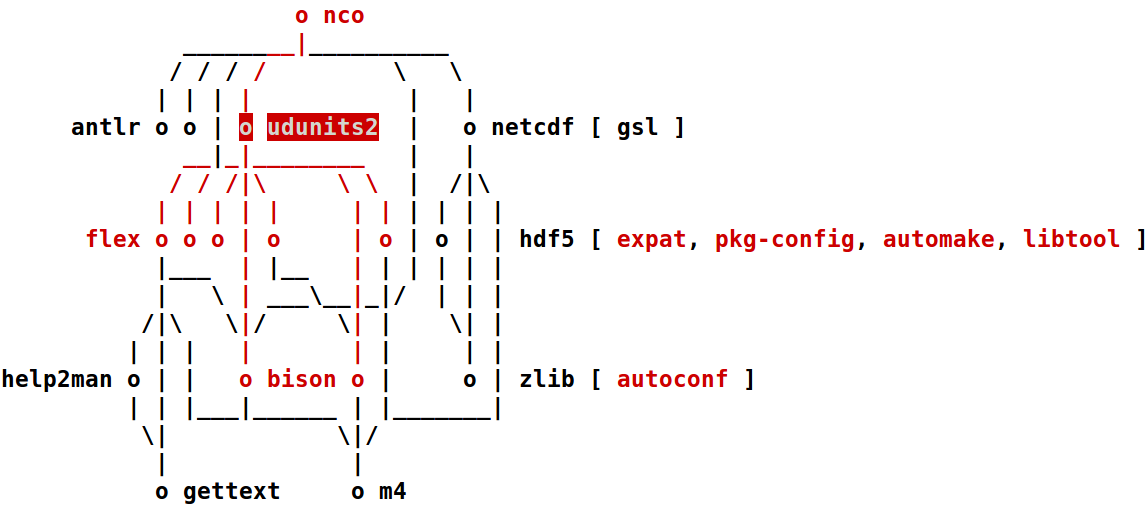}
}

\subfloat[All connected nodes and paths highlighted.\label{fig:nco:reachability}]{
    \includegraphics[width=1.0\columnwidth]{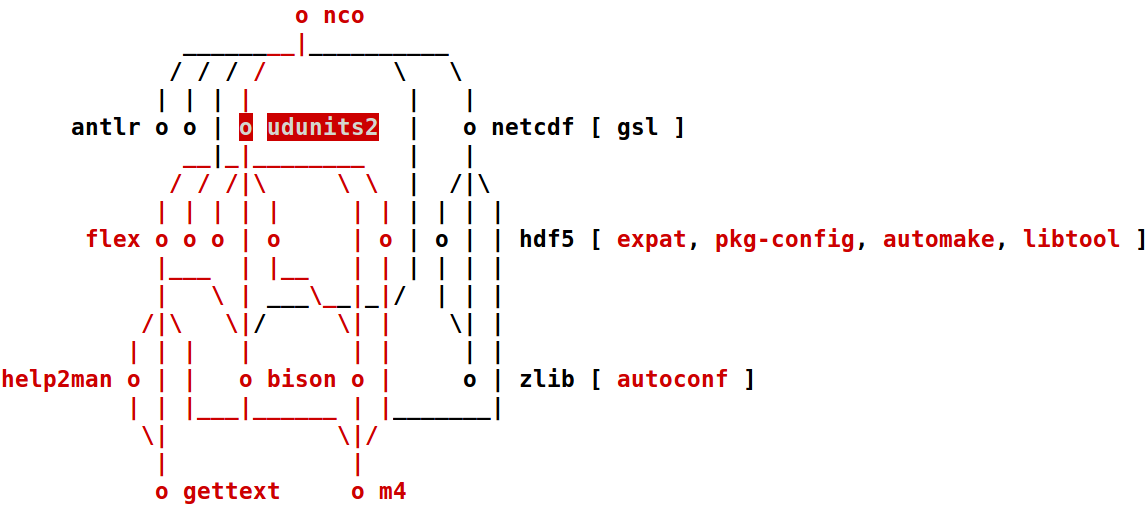}
}
    \caption{Two highlighting styles: (a) only
    direct connections (single edge paths), and (b) showing all connected nodes
    (multi-edge paths included). In (b), \texttt{bison}, \texttt{m4},
    \texttt{help2man} and the edges to them are highlighted in addition to
    \texttt{udunits2}'s
    direct neighbors.}
    \vspace{-0.05in}
\label{fig:nco}
\end{figure}

Users may traverse the nodes (via highlight) in grid order with the \texttt{n}
and \texttt{p} keys, similar to jumping between found matching strings in
\texttt{less}. These interactions support examining multiple nodes or
gaining and understanding of the edge coalescing for graph overview
tasks.

With the exception of the arrow keys, which are not available in all
terminals, we did not find consensus for directional movement. Thus, we
provide both arrow keys and the set \{w, a, s, d\} for panning should the
graph not fit in the viewable area of the terminal. The latter set was chosen
for its prevalence for inputting directional movement in video games.
Searching for a node also automatically pans the graph to ensure the node
is on screen.  Zooming is possible on some terminals by changing the font
size.

In addition to helping users disambiguate bundled edges and associate nodes
with labels, we expect the automatic highlighting of connected edges and
neighbors to a node (as in Fig.~\ref{fig:teaser}, Fig.~\ref{fig:mkfontdir:term},
Fig.~\ref{fig:nco:direct}) to help with connectivity and accessibility tasks
like those described in Sec.~\ref{sec:taskabstraction}.  Highlighting has been
shown to aid users in these visual graph queries~\cite{Ware2004}. Users may
toggle the highlighting to highlight all nodes with a path to or from the
highlighted node instead, along with the edges in those paths
(Fig.~\ref{fig:nco:reachability}). 

In interactive mode, \texttt{graphterm} exploits the entirety of the terminal
window via the terminal-independent \texttt{ncurses} text interface library.
Upon quitting interactive mode, the graph is printed to the terminal in the
state it was last shown, with the exception that all rows are printed rather
than only what would fit in the terminal's display.

\section{Study}
\label{sec:study}
Our goal in designing \texttt{graphterm} is to provide dependency graph
visualization in the context in which people need to analyze them, at the
command line, so as to benefit their workflow. In our discussions, Spack
maintainers also expressed a strong preference for a visualization that could
be used at a terminal. However, as discussed in Sec.~\ref{sec:git} and
Sec.~\ref{sec:ascii}, the ASCII representations have limitations not present in
the graphical space. We conduct a study to observe the efficiency trade-offs
between the visualizations as well as user preferences when visualizing
dependency graphs during a command-line workflow.

Sensalire et al.~\cite{Sensalire2009} proposed a taxonomy of software
visualizatin tool evaluations. The taxonomy consists of ten dimensions
(S1--S10): (S1) tool selection, (S2) participant, (S3) tool exposure, (S4)
task selection, (S5) experiment duration, (S6) experiment location, (S7)
experiment input, (S8) participant motivation, (S9) participant relation to
tool designers, and (S10) analysis of results.  We note the relation of our
design to thise taxonomy throughout.

We design a within-subjects study comparing the \texttt{graphterm},
\texttt{git}-like, and GraphViz dependency graph visualizations. We
hypothesized:
\begin{enumerate}[label=H\arabic*. ]
\item Participants who use command line interfaces will
prefer ASCII representations.

\item Participants who use command line interfaces will perform
tasks more quickly using ASCII representations when working at the command
    line.

\item There will be no significant difference in accuracy among the
  visualizations, but participants will be more confident in their answers
    when using GraphViz.
\end{enumerate}

\noindent\textbf{Tool Selection (S1).}
We chose GraphViz for comparison because it is widely used, can generate
image files from the command line, and is already suggested by Spack.  In
addition to the \texttt{git}-like ASCII graphs, Spack can output a graph in
\texttt{dot} file format. The Spack documentation suggests its use with
GraphViz to generate a PDF file. 

Koschke~\cite{Koschke2003} surveyed software engineers and found their most
commonly used graph layout program was \texttt{dot}. Our survey of Github
repositories (Sec.~\ref{sec:related}) showed the most common approach for
generating a visualization of dependencies was outputting a \texttt{dot} format file for
rendering with GraphViz. We conclude from these findings that Spack's use of
\texttt{dot} and GraphViz is in-line with current practices and therefore it
is appropriate to compare this GraphViz-based workflow to the ASCII
visualization workflows. For our study, we preserved the command line
suggested by Spack, which was similar to those suggested by many of the Github
projects.

Spack's suggestion of rendering \texttt{dot} to PDF (rather than to PNG) was
also kept because PDF readers have built-in functionality for searching and
highlighting text, such as the label of a node in the graph, which while not
required by the tasks, could be used to help locate nodes specified by the
questions. 

We kept all of the graph style attributes written into the \texttt{dot} file
by Spack. We removed the hash ID from the node labels however as these might
confuse participants unfamiliar with Spack.

\begin{figure}
    \centering
    \includegraphics[width=1.0\columnwidth]{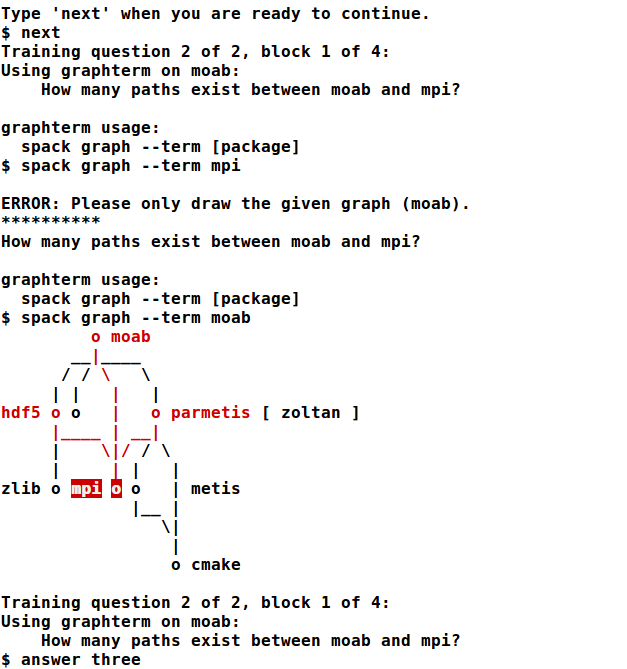}
    \caption{Study questions, answers, and operations all take place at the
      command line. Part of the \texttt{graphterm} training phase is
    shown here.}
    \label{fig:moab-study}
    \vspace{-0.05in}
\end{figure}

As we want to investigate the utility of the package dependency graph
visualizations for the {\em workflow} of people using the command line, it was
essential that the study be conducted at a command line interface. Our study
software displays instructions, tutorials, and questions on the terminal and
accepts answers via terminal commands. Fig.~\ref{fig:moab-study} shows the
interface. Participants were allowed to use the command line as they wished,
with the exception that our software prevented them from visualizing the wrong
graph or using a disallowed method (e.g. using \texttt{graphterm} during the
GraphViz block). This limitation applied only to the \texttt{spack graph}
command---participants could render \texttt{dot} files to PNG or look at the
raw text format if they chose to do so. Similarly, as we wanted to better
emulate real-world Spack workflow, we used Spack to generate the graphs rather
than relying on pre-generated graphs.

To measure the efficacy of the {\em workflow}, we chose to measure response
time in terms of the entirety of the workflow---from question to response, this
allowed the user to perform other operations as they wished, as they would be
able to in a real setting.


\vspace{0.5ex}

\noindent
\textbf{Study Design}. Our study is divided into four blocks, one for each
visualization workflow (`tool blocks') and an additional final block where participants
could use the visualization of their choice. The three tool blocks contained
five questions each while the user-preference block contained three questions.
General command lines (``usage'') were given with each question.  In the
user-preference block, the questions were prefaced with the phrase `Using any
method' and command lines for all three methods were displayed.

The order of the tool blocks and the question order within them was randomized. A brief tutorial describing package dependency
graphs was displayed at the beginning of the study. Before each tool block, an
explanation of the particular visualization scheme was given, two interactive sample
questions with answers were presented, and participants were encouraged to
experiment with the tools and given a list of package names with which to do
so.
At the end of the study, the software asked the
participants the open-ended questions:
\begin{itemize}
  \itemsep=1ex
  \item ``Which graph type do you prefer and why?''
  \item ``What features made the graphs hard or easy to understand?''
\end{itemize}
The final question solicited any additional comments. Participants required
between 45 and 65 minutes of active time to complete the experiment (S5: Experiment Duration).

\vspace{0.5ex}

\noindent\textbf{Task Selection (S4).}
Participants were asked two types of questions based on the tasks determined
in \ref{sec:background}: direct connectivity and path counting. The direct
connectivity questions ask which packages depend on a given package in the
graph ({\em affected by} connectivity). This operation occurs when a user is determining which compiler or
version to set for a package and how that may affect the packages that depend
on it. The path counting questions ask how many paths exist between two nodes
of a given graph. This operation can be helpful for debugging or developing
new features for the package management system as maintainers try to
understand why the system resolved dependencies in a
particular way. All questions were open-response to better match real world
workflow. We decided against multiple choice questions as we were concerned
they would enable process of
elimination and guessing.

\vspace{0.5ex}

\noindent\textbf{Experiment Input (S7).}
We used different graphs for all four blocks to avoid memory effects. To
balance the difficulty across the three tool blocks, we chose graphs with
similar vertex counts, edge counts, layers, and nodes per layer.
Table~\ref{tab:studygraphs} summarizes the chosen graphs for the tool blocks as
well as the preference block\footnote{See appendix for further
descriptions of the graphs and questions.}. The type of question used for each
is also listed. Note for the largest graph size, we asked both types of
question for the same graph due to the limited selection at that size. To
further balance difficulty, we created questions targeting similar layers in each
graph with similar answers in terms of numbers of nodes or paths. We chose
several graphs per block to test a variety of graphs without incurring tedium
and fatigue, which was reported by participants during piloting. All graphs
were generated with the \texttt{--normalize} option in Spack to ensure the
same structure across different systems.

\vspace{0.5ex}

\noindent\textbf{Experiment Location (S6).}
To reach a wide audience of participants, we provided both in-person and
online options for accessing the study. There were three online options: (1)
loading files on a system on which many Spack users have shell access, (2)
running the study within a provided virtual machine, or (3) running the study
in a Docker~\cite{Merkel2014Docker} container. The in-person and virtual
machine versions had default terminal configurations that were screen height,
had black backgrounds, and had all other settings left with the system default
(e.g, font sizes). We did not do any special configuration on the
visualization programs or any PDF or image viewer. Participants were free to
alter configurations such as window size and zoom level at any time, including
during the training phase. 

\begin{table}[h]
  \caption{Dependency Graph Characteristics for Study}
  \label{tab:studygraphs}
  \scriptsize
  \centering
  \begin{tabu}{
      l%
      *{2}{c}%
      *{1}{l}%
    }
  \toprule
    Block(s) & \# Nodes & \# Edges & Question Type \\
  \midrule
    Tool & 11 & 22 & Paths \\
  \midrule
    Tool & 17-18 & 26-27 & Dependencies \\ 
  \midrule
    Tool & 22 & 45 & Paths \\
  \midrule
    Tool & 33-34 & 60-62 & Paths, Dependencies \\
  \midrule
    Preference & 19 & 27 & Paths \\
  \midrule 
    Preference & 31 & 62 & Dependencies \\
  \midrule
    Preference & 40 & 79 & Paths \\
  \bottomrule
  \end{tabu}
\end{table}

We verified the time each visualization required was approximately equal
between graphs.  We measured the time to produce a layout for each
experimental object (in random order) seven times on the machine used for the
in-person sessions. We chose to compare median time as we observed the
distribution of the seven timings was usually either a few milliseconds
different or an entire second different.  The GraphViz time included the time
to render to PDF, but not to open the PDF as we allowed participants to use
any viewer they chose. The \texttt{graphterm} time was measured using printing
to \texttt{stdout} instead of interactive mode. All experimental objects of
approximately the same node count took the same time (in rounded milliseconds)
with the exception of the 40 node Preference block graph and the 33-34 node
Tool block graph where GraphViz took 1 second longer (three trials took the
same time as the other visualizations, four took one second longer). 

The version of \texttt{graphterm} used in the study relied on the Tulip Python
bindings. The self-contained version currently on Github ports the subset of
the Tulip hierarchical layout used by \texttt{graphterm}.  This Python-only
version is significantly faster both because of the reduced algorithm and
not having to connect to the Tulip interface. Our measurements were done
using the study version.

\vspace{0.5ex}

\noindent 
\textbf{Participants (S2, S8, S9) and Tool Exposure (S3)}. We recruited participants from both the Spack
community (including maintainers, package developers, and users) and our
organizations' computing departments. Nineteen people completed the study.
We asked participants about their command line usage during the initial questions
of our study. Of them, we discarded any results where the participant could
not answer the majority of questions in any of the three tool blocks correctly
(two participants), where the participant indicated they never use the command
line (one participant), or where the participant indicated they did not
understand what a path was after training (one participant). This left us with
15 responses (ten men and five women; eight between 18-25 years of age and
seven between 26-45 years of age). Six people participated remotely and nine
in person. Three of the participants (S1, S2, S3) were Spack users. S1 was
also a Spack contributor. S2 indicated they had written small Spack packages.

None of the participants were involved in the analysis or development
presented in this paper. None had seen or used \texttt{graphterm} before. Of
the 15 participants, 7 had used GraphViz before and 14 had used \texttt{git}.
Both groups included all the Spack users. We did not ask whether the
participants were familiar with the \texttt{git log graph} command or what
their motivation for participating was.

\section{Results and Analysis}
\label{sec:analysis}
\begin{figure*}[t]
    \centering
    \includegraphics[width=1.0\textwidth]{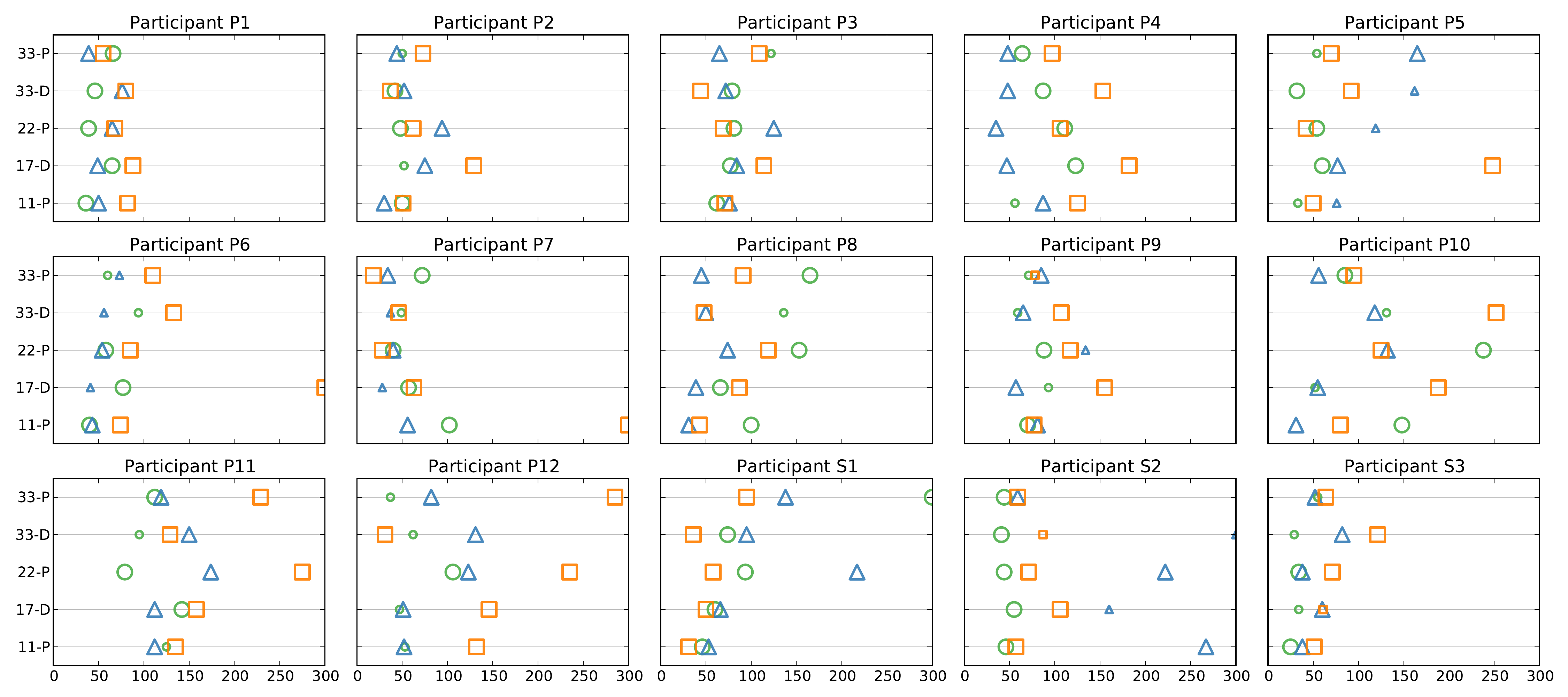}
    \includegraphics[width=1.0\textwidth]{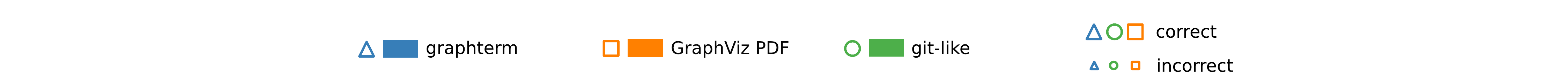}
\caption{Response time and accuracy results for study participants during the tool
  blocks, organized by graph size and question type (path P or dependencies
  D). Time is shown in seconds on the horizontal axis. Response time includes
  visualization rendering time. 
  The blue triangles are \texttt{graphterm}-aided responses, the green
circles are responses using the \texttt{git}-like depiction, and the orange
  squares are responses with the
GraphViz-rendered PDFs. Smaller marks indicate incorrect responses. Response times greater than 300
seconds are clamped to the plot edge and were not included in the statistical
analysis.}
    \vspace{-0.05in}
\label{fig:studyresults}
\end{figure*}

\begin{figure*}[t]
    \centering
    \includegraphics[width=1.0\textwidth]{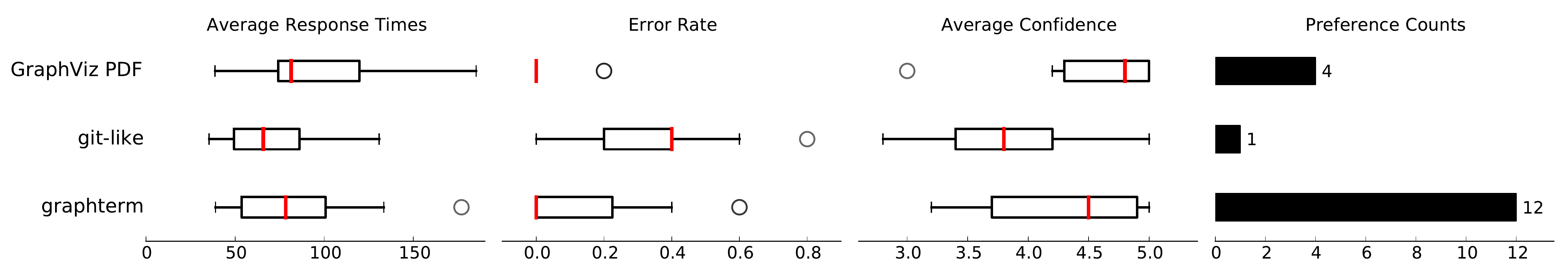}
\caption{Summary plots of participant average response time, error rate, and
  confidence scores along with reported preferences. The error rate boxplot
  has overlapping outliers (3 at 0.2 for GraphViz, 1 at 0.8 for the
  \texttt{git}-like, and 2 at 0.6 for \texttt{graphterm}).
  Preference values sum to
  greater than 15 as some participants reported multiple preferred workflows.
  }
    \vspace{-0.05in}
\label{fig:boxplots}
\end{figure*}

We analyze the data collected during our study. As the goal of our study was
to assess efficiency and preference (S10: Analysis of Results) in comparison
to existing approaches, we assess response
time, accuracy, and confidence across the three visualizations using a linear
mixed effects model. We then examine the preferences of our participants as
expressed directly and through tasks. We identify and review common themes
communicated by the participants about the visualizations. Finally, we discuss
our conclusions based on these results as well as limitations of the study.

Fig.~\ref{fig:studyresults} shows the measured response time and correctness
for the tool block responses of each participant, organized by the graph size
and question type. Fig.~\ref{fig:boxplots} summarizes the spread of average
participant response time, error rate, and confidence scores and shows
preferences reported by participants.

We measured response time from when the question was shown through when the
answer was entered.  Based on our pilot, we omitted any single question time
longer than five minutes (shown clamped at 300 seconds in
Fig.~\ref{fig:studyresults}). This is generally an indication the participant
took a break from the study. In total, we removed four question responses all
from different users: two using GraphViz PDFs, one using the \texttt{git}-like
visualization, and one using \texttt{graphterm}. In one response, the
participant indicated they had left and returned, so we updated the time
recording as noted. In scoring the answers, we accepted package names with
typos (e.g., `xrpoto' instead of `xproto') as all instances were unambiguous
with respect to other names in the graph.  

To account for within-subjects variance and missing trials, we built a linear
mixed effects model, previously used in the visualization community by Liu and
Heer~\cite{Liu2014}. We model the visualization used and whether the sample
was collected in-person or remotely as fixed effects. We model the size of the
graph in question (node count) as a random intercept term. As we expect
different reactions to the visualization by participant, we model participant
as a random slope term modulated by the visualization used. We assess significance with
a likelihood-ratio test~\cite{Winter2013} using a reduced model without the
visualization term.  Post-hoc analysis on significant findings is performed
using Tukey's all-pairs comparisons with a Bonferroni-Holm correction.  For
these analyses, we use the R~\cite{R} \texttt{lme4}~\cite{lme4Paper} package
for the model and the \texttt{multcomp}~\cite{multcomp} package for post-hoc
analysis of significant findings.

\vspace{1ex}

\noindent
\textbf{ASCII Visualizations Resulted in Decreased Response Times}. We found a
weak effect for visualization method on task completion time, $\chi^2(2, N=15)
= 5.812$, $p < 0.1$, with summary coefficients of -26.89 seconds for the
\texttt{git}-like visualization workflow and -12.59 seconds for the
\texttt{graphterm} workflow,
indicating tasks were completed more quickly in the ASCII visualization
workflows.
Post-hoc analysis showed significance in the difference between
the \texttt{git}-like visualization and GraphViz PDF visualization ($p <
0.05$). This finding partially supports H2, that participants would perform
tasks more quickly with ASCII-based workflows when working at the command
line.

Analyzing the response logs, we noted Participant S2 wrote detailed comments
with their answers in the \texttt{graphterm} block, increasing the time
recorded for response in that block. Also, some participants realized that as
one graph was the subject of two questions, they need not re-create the
GraphViz PDF. They simply re-opened it which resulted in a savings of Spack
dependency resolution time, graph rendering time, and possible memory effects.

\vspace{1ex}

\noindent \textbf{ASCII Visualizations Resulted in More Errors}.  We found a
significant effect for visualization method on incorrect responses $\chi^2(2,
N=15) = 6.8848$, $p < 0.05$ with summary coefficients of -0.206 for the
\texttt{git}-like visualization workflow and -0.094 for the \texttt{graphterm}
workflow, indicating more errors occurred for the ASCII-based workflows.
Correctness was recorded as a binary value with 1 for correct and 0 for
incorrect. Post-hoc analysis showed a significant difference in errors between
the \texttt{git}-like method and the GraphViz PDF method ($p
< 0.05$). This finding is contrary to the first part of H3, that there would
not be a significant difference in accuracy among the workflows.

We analyzed the response logs for trends and explanations. Participant P5 and
P6's answers in both ASCII visualizations suggest they did not understand the
implied edge direction by orientation. Participants P5 and P6 made six of the total ten
errors in the \texttt{graphterm} worfklow and four of the 23 total
errors in the \texttt{git}-like workflow. Participant P7 was randomly
shown several path counting questions in a row during the \texttt{graphterm}
block.  The dependency questions afterward were answered with the (correct)
number of dependencies rather than the names and thus were marked incorrect.
Participant P10 made similar mistakes in the \texttt{git}-like block. These
mistakes represented two of the ten errors made with \texttt{graphterm}
workflow and two of the 23 errors made with the \texttt{git}-like workflow. In
contrast, only three errors were made using the GraphViz PDF workflow.

\vspace{0.5ex}

\noindent
\textbf{ASCII Visualizations Resulted in Lower Response Confidence}.
We also found a significant effect for visualization method on response
confidence $\chi^2(2, N=15) = 7.163$, $p < 0.5$ with summary coefficients of -0.607 for
the \texttt{git}-like visualization workflow and -0.283 for the \texttt{graphterm}
workflow, indicating participants were less confident in their answers
using ASCII visualizations. Participants were asked to rate their confidence on a scale of
1 to 5 after each question. Post-hoc analysis showed a significant difference
between the \texttt{git}-like visualization workflow and the GraphViz PDF
workflow
($p < 0.01$). This finding supports the second part of H3, that participants
would be more confident when using the GraphViz PDFs.

\vspace{1ex}

\noindent
\textbf{\texttt{graphterm} Was Preferred By Participants}.
During the preference block, eleven of the fifteen participants used
\texttt{graphterm} most frequently, with nine using only \texttt{graphterm},
including all participants with Spack experience (S1, S2, S3).  Two used the
GraphViz PDFs exclusively, one used the \texttt{git}-like and GraphViz PDFs
equally, and one accessed graphs in a way our system was unable to record
(this participant reported preferring \texttt{graphterm}).  

When asked directly at the end of the study, eleven of the fifteen
participants indicated that they preferred \texttt{graphterm} to the other two
visualizations, again including all participants with Spack experience.
Participant S3 wrote ``after using the \texttt{--term option}, I doubt I will
ever use any other graph format except when printing it out: for printing,
\texttt{--dot} is better.'' Of the other four participants, one indicated they
liked both \texttt{graphterm} and GraphViz PDFs, another liked both the
\texttt{git}-like visualization and GraphViz PDFs, and two preferred only the
GraphViz PDFs. One of the participants who preferred the GraphViz PDFs issued
the caveat ``if I had more practice with the term graphs, I feel as if that
would be more efficient and clear.''

The results of the preference block and the preference question partially
support H1, that participants would prefer the ASCII representations. While
participants showed preferences for \texttt{graphterm}, preference towards the
\texttt{git}-like representation was minimal.

Several participants answered questions in the preference block correctly,
even when they had shown less accuracy with the chosen drawing method in its
initial block (S2, P7, P9). While P7 can be attributed to misreading the
earlier questions, the increased accuracy for S2 and P9 may indicate learning
over the course of the study. 

\subsection{Themes in Participant Responses}
\label{sec:qual}

We discuss common themes found in comments made by the participants, both in
response to the open-ended questions at the end of the study and any comments
typed during the study in their answers.

\vspace{0.5ex}

\noindent
\textbf{Use of Space}. 
Participant S2 stated an important
factor was the ``use of screen space both vertically and horizontally,''
noting there is more horizontal screen space to spare. Participant P7 preferred
\texttt{graphterm} for its ``good use of horizontal spacing.''
Participant P2 thought the aspect ratio of the \texttt{git}-like visualization
was detrimental. 

The rank-based nature of all the layouts was considered useful by Participant
S1, who noted ``The top-to-bottom direction also really simplifies things, in
all the graphs.''

\vspace{0.5ex}

\noindent
\textbf{Scrolling}. Many participants said scrolling was detrimental. This
was most prevalent for the GraphViz PDF visualization (S1, S2, P2, P4, P5, P11),
but also for the \texttt{git}-like visualization (S1, P4).
Participant S1 remarked ``its tedious to find nodes and move around the pdf.''

\vspace{0.5ex}

\noindent
\textbf{Ambiguity}. Several participants expressed some difficulty with the
ambiguity of edge connectivity in the \texttt{graphterm} layout (S1, S2, P8),
but said the highlighting helped (S1, S2). Some participants noted that
understanding edge connectivity was also tricky in the GraphViz PDF (S1, S2,
S3, P4) due to crossings, but Participant P11 indicated that it was easier as
no edges branched like they do in both the \texttt{git}-like visualization and
\texttt{graphterm}. While Participant P6 preferred \texttt{graphterm} and
GraphViz PDF to the \texttt{git}-like visualization, they stated none of the
tools were easy to understand.

\vspace{0.5ex}

\noindent
\textbf{\texttt{git}-like visualization}. Several participants indicated the edge
coloring helped them trace paths in the \texttt{git}-like visualization (S1,
S2, P2, P4, P5), but two experienced difficulty discerning some of the colors (P8,
P11). The density of the \texttt{git}-like visualization was considered a
negative (S2, P2, P5, P11). Participant S2 wrote ``my initial reaction to the
large git graphs was a viscrecal [sic] - I do not want to look at this at
all.'' 

\vspace{0.5ex}

\noindent
\textbf{GraphViz PDFs}. Two of the participants who preferred the GraphViz
PDFs (P3, P5) said that the arrowheads clarified what the dependencies were.
Participant P3 also said the direct labeling of nodes was helpful. Though
participant P11 preferred \texttt{graphterm}, they also remarked they liked
the arrows in the GraphViz PDFs. 

Three participants expressed distaste for using a PDF reader (S1, S2, P9).
Participant S2 wrote that ``having to load pdfs is annoying.'' While some
participants during piloting reported using their PDF reader's text search to
locate a node, this was not reported during our full study. Instead,
participants described having difficulty finding a node in the GraphViz
layout (S1, S3, P10). Some participants suggested modifying the default
GraphViz PDF rendering to have bigger fonts (P8) and shorter edges (S2, P11).

\vspace{0.5ex}

\noindent
\textbf{\texttt{graphterm}}. Many participants cited the interactive
highlighting of \texttt{graphterm} as a key feature (S1, S2, P2, P6, P7,
P8, P9, P11, P12). Two participants (S1, P11) suggested enhancing the
highlighting modes to color differently for direction or extend by
neighborhood. 

As expected, the distance between the node and the label was found confusing
by participants (S2, P11). Participant P11 wrote ``The fact that multiple
node names on the same line get grouped together is annoying.''

Some participants indicated the use of the terminal as another reason for
their preference (S1, S3, P9).  Participant P9 wrote that \texttt{graphterm}
``was just RIGHT there, straightforward to use especially when you just want
to use the command line the whole time.'' Participant S3 disliked that
GraphViz ``is not terminal based'' and liked \texttt{graphterm}'s ``keyboard
based navigation.''

\subsection{Discussion}
\label{sec:discussion}

Based on our quantitative measures, we observe that the workflow using the
\texttt{git}-like ASCII visualization leads to faster response times from the
command line than workflow using the GraphViz PDFs, but at a cost to accuracy
and confidence. While not statistically significant, the workflow with
\texttt{graphterm} seems to fall between the two existing Spack dependency
graph visualizations on these three measures, from which we infer the
\texttt{graphterm} workflow is a viable alternative to the existing Spack
graph offerings.

Accuracy is a significant concern when making build decisions. We note none of
the three options were strictly error-free. While the GraphViz workflow
resulted in three errors total and the \texttt{graphterm} workflow ten, two of
the \texttt{graphterm} errors can be attributed to misread questions and six
to insufficient training (see Sec.~\ref{sec:validity} below), indicating the
error rate in practice may be comparable.

The GraphViz rendered visualizations can directly label the nodes unlike
\texttt{graphterm} and are less ambiguous. Yet, the workflow with the
\texttt{graphterm} visualization was more preferred.  Based on the comments by
participants, we believe that the major factors leading to this preference
were the terminal-based nature and the interactivity. We interpret the
preference for the terminal (when already working at the terminal) to indicate
participants are willing to accept a sub-optimal visualization that is
convenient to their workflow. However, as the workflow with the
\texttt{git}-like graphs were not preferred, this trade-off between
visualization and workflow is not absolute. A graphical solution with more
customized interactivity may be preferable to all three presented options.
When we suggested such a solution to users during our task analysis
(Sec.~\ref{sec:taskabstraction}), they indicated having something at the
command line was of greater interest. The preference results are in line with
the initial assessment of the domain experts.

Some of the issues participants noted in the GraphViz PDFs could be addressed
by having Spack change the graph layout style attributes it writes into the
\texttt{dot} file, such as the font size and edge length changes proposed by
the participants. Both of these parameters were already explicitly written by
Spack.

\subsection{Limitations}
\label{sec:validity}

Our findings are limited by the study design as described below.

\vspace{1ex}

\noindent\textbf{Task Design and Study Length}.
We designed our study questions to test basic graph tasks derived from our
task analysis (Sec.~\ref{sec:taskabstraction}) rather than the more complicated
task a real user may have. The more complicated tasks often rely on
familiarity with software, its options (e.g., available parallel runtime implementations),
and personal taste of the user (e.g., favored compiler). Some, like debugging
Spack itself, require knowledge of the Spack codebase. We expect the more
complicated tasks will involve several basic graph tasks per dependency graph.
The cost of obtaining a graphical representation may be amortized over this
process. Alternatively, the barriers to interacting with such a
representation, e.g. switching between the terminal and another program, may
compound.

Furthermore, while our goal was to examine usage coming from a command line
workflow, the repeated visualization of different graphs in sequence likely
does not match the workflow of Spack users, who probably visualize graphs more
infrequently. Another study which spaces out visualization with other command
line tasks may better emulate reality, but would increase the length of the
study.

Study length was likely a factor in our ability to recruit Spack users as
participants, though it was half the maximum time of two hours suggested by
Sensalire et al.~\cite{Sensalire2009} when recruiting professionals. A
follow-up study could bypass the graph layout time by pre-computing the graphs
as these operations were equal across all layouts, at the cost of realism in
the presented workflow. 

\vspace{1ex}

\noindent\textbf{Effect of Study Setup on Response Time}.
Participants either performed the study on a local machine (11 participants)
or were warned ahead of time about viewing multiple GraphViz-rendered files
such that they might want to access the study with X11 forwarding enabled
(four participants). Participants did not experience the scenario where a
separate login operation or file copy was required to view a PDF or image
file.  Avoiding these scenarios by either working locally or being informed
ahead of time may have resulted in quicker task response time when using the
GraphViz PDFs. 

We used GraphViz in our comparison as it is one of the existing options
supported and suggested by Spack, it is widely used in the systems and
software space, and visualizations can be generated from the command line in
relatively few steps and without users having to learn new technology.  We
used the \texttt{dot} specification as provided by Spack. A customized
graphical visualization or more style specification in the \texttt{dot} file
may lead to better performance. Reminding participants that text search is a
common feature in PDFs may have helped participants who stated they had
difficulty finding nodes. Fully integrating the file copy and graphical
viewer launch may also lead to better performance of GraphViz, but would require assumptions that
limit portability of Spack or place a per-machine configuration burden on the
user. This would increase the cost of the whole of the workflow, especially for
users who consult the graphs infrequently and would not 
be amortizing the setup cost.

\vspace{1ex}

\noindent\textbf{Effect of Study Setup on Error Rate}.
The majority of the errors in the \texttt{graphterm} workflow and several in
the \texttt{git}-like depiction workflow came from participants P5 and P6. The
responses were indicative of not understanding that edge directionality was
implied by vertical positioning, despite being explained in the training phase
of both ASCII workflows. Participants S2 and
P9 who made errors during the \texttt{graphterm} block did not make the same
errors using the \texttt{graphterm} workflow in the preference block, which
may indicate improvement over the course of the study. 
These observations may indicate the training was insufficient, leading to
increased errors for the ASCII workflows.

\vspace{1ex}

\noindent\textbf{Effect of Free Response Questions}.
We chose free response questions over multiple choice to better match a
realistic scenario and to
avoid guess-and-check
behavior. This allowed participants to answer questions other than what was
posed (e.g., answering the number of packages instead of their names),
increasing the error rate for the ASCII workflows. It also lead to increased
response times for participant S2 during the \texttt{graphterm} block as this
participant wrote comments with their answers.

\section{Conclusion and Future Work}
\label{sec:conclusion}
We have described and validated a visual solution for analyzing Spack package
dependency graphs for command-line workflows. The solution combines ASCII
characters with terminal interactivity to provide DAG visualization that
supports topology-centric graph tasks identified in our task analysis and
abstraction (\autoref{sec:taskabstraction}). Our DAG visualization software,
\texttt{graphterm}, provides a terminal-friendly compact depiction by adapting
an existing graphical layout algorithm. These adaptations include selectively
re-routing and bundling edges and choice of ASCII marks. The interactivity
leverages idioms from widely-used command line tools.

We designed a user study to compare the analysis workflow using
\texttt{graphterm} to the workflows using the two existing package dependency
visualizations for two of our identified tasks. The study emulated the setting
in which Spack users work: the command line interface. Participants in our
study preferred the \texttt{graphterm} workflow to the existing approaches.
Based on the responses to this study, we conclude that though ASCII-based
visualization requires trade-offs not present in the graphical space, it can
be a more preferable visualization solution from the perspective of the entire
workflow. We hope this demonstration leads to more consideration of making
these stark trade-offs to support the analysis of users where they are (in our
case, the terminal) and note there is a lot of opportunity for interactive
tools in this space.  

Having shown the viability of visualizing small graphs in ASCII, we intend to
further improve the layered ASCII graph layout algorithm in the future.  We
focused on bundling and re-routing diagonal edges for clarity and to make the
graph less tall. We would like to further explore the affect of the bundling
on readability, possibly also considering bundling and re-routing vertical
edges to make the graph less wide. Additionally, we plan to explore shifting
nodes so there are fewer per line, considering unique marks for the nodes, or
adding tooltip-like support to address issues with node labeling. 

\texttt{graphterm} was designed with an emphasis on graph topology tasks that
we identified as common across the users of package dependency graphs.
However, we also identified an interest in graph attribute data among Spack
developers. While some of this information is available through the indented
trees, no attempt has been made to address this in the existing graph
features. We plan to address the multivariate design issues in the ASCII
space. We will also experiment with adding interactivity to other Spack
analysis commands or implementing support for multiple coordinate views to
further support these tasks.

\section{Acknowledgements}
The authors thank the Spack community, study participants, and reviewers for
their time and aid in this research.  This work was performed under the
auspices of the U.S. Department of Energy by Lawrence Livermore National
Laboratory under Contract DE-AC52-07NA27344. LLNL-JRNL-746358.

\bibliographystyle{abbrv-doi}
\bibliography{paper}

\appendix

\section{Experimental Objects used in Study}

\begin{table*}[h]
  \caption{Dependency Graph Characteristics for Tool Blocks}
  \label{tab:studygraphsplusplus}
  \scriptsize
  \centering
  \begin{tabu}{
      *{7}{l}%
    }
  \toprule
    Tool & \# Nodes & \# Edges & Layers & Nodes per Layer & Question & Question Layers \\
  \midrule
    GraphViz & 11 & 22 & 4 & 1-2-5-3 & paths & 1 \& 3\\
    \texttt{git}-like & 11 & 22 & 4 & 1-2-5-3 & paths & 1 \& 4 \\
    \texttt{graphterm} & 11 & 22 & 5 & 1-1-2-4-3 & paths & 1 \& 4\\
  \midrule
    GraphViz & 17 & 27 & 6 & 1-1-4-2-6-3 & dependencies & 5\\
    \texttt{git}-like & 17 & 26 & 7 & 1-2-4-2-1-4-3 & dependencies & 4 \\
    \texttt{graphterm} & 18 & 26 & 5 & 1-4-6-5-2 & dependencies & 4\\
  \midrule  
    GraphViz & 22 & 45 & 7 & 1-1-1-4-6-6-3 & paths & 1 \& 5 \\
    \texttt{git}-like & 22 & 45 & 7 & 1-1-1-4-6-6-3 & paths & 1 \& 5\\
    \texttt{graphterm} & 22 & 45 & 7 & 1-1-1-4-6-6-3 & paths & 1 \& 5\\
  \midrule
    GraphViz & 34 & 62 & 7 & 1-2-4-5-5-11-5-1 & dependencies & 8 \\
    & & & & & paths & 1 \& 7 \\
    \texttt{git}-like & 33 & 60 & 7 & 1-1-4-5-5-11-5-1 & dependencies & 6 \\
    & & & & & paths & 1 \& 7 \\
    \texttt{graphterm} & 34 & 61 & 7 & 1-2-4-5-5-11-5-1 & dependencies & 6\\
    & & & & & paths & 1 \& 7 \\
  \bottomrule
  \end{tabu}
\end{table*}

We provide more details about the experimental objects used in the study.
Table~\ref{tab:studygraphsplusplus} appends information about the number of
layers per graph, the division of nodes in each layer, and the layers on which
the target nodes for each question resided for the Tool block graphs. In some
cases we were able to choose graphs that had similar major dependencies,
resulting in highly similar graph structures.

\section{Dependency Visualization Features in Github Repositories}

We performed a online search for existing methods for visualizing dependencies
using the search string ``site:github.com visualize dependencies.'' We used
Google search in an incognito tab of a newly installed Google Chrome with no
signed in accounts.  We manually retrieved links from the 49 returned pages.
Some links were duplicated between results pages. We added additional links
found from examining the total ones (procedure described below), resulting in
a total of 521 links examined. 

We explored each link to determine if it had a dependency visualization
feature. If the returned link was not a project page (e.g., it was an issue,
feature request, or project file) and did not mention visualization, we
navigated to the main project page of the returned link. If the returned link
referred to a project we already analyzed, we skipped it. We also skipped
projects that were general visualization tools or were not targeted at a
computing domain.  For example, we skipped all projects where the target of
the visualization was biology-related. We included machine learning network
visualizations but not sentence structure dependencies in natural language
processing or scene graph overlays in computer vision.  We skipped links that
were blog posts or personal websites. We manually inspected stand-alone code
snippets (`gists').  If the returned link was a discussion thread (e.g., for
an issue or feature request) and mentioned a possible outside-project
visualization, we followed those links. For projects that were lists of links
to other projects, we searched for promising links using the strings `visual'
and `dependenc.'  Of the 521 links, we found 483 unique projects.

From the main page of any project, we assessed the graph visualization
features in the following manner. If we could discern enough information from
the link returned by the search (sometimes the manual or wiki) or the README,
we associated the found features with the project. If the graph visualization
procedure was not fully explained (e.g., an image was shown but the libraries
to generate it were not described or text contained something like `visualizes
with GraphViz'), we performed a directed search of the code to determine what
the partial explanation meant. If no visualization procedure was mentioned in
the README, we either read the entirety of the source code (for small code
bases) or directed our search using Github to search the repository for the
terms: (graph, network, tree, visual, view, plot, diagram, layout, svg, png,
pdf, html, dot, gexf, graphml, dagre, d3, indent, ascii). 

For the keyword search, we used all terms even after finding a visualization
so that no keyword or visualization would get preference simply due to order
of the search. We manually inspected the snippets returned in
the first five pages of results for each term if they existed and searched
further on promising leads.  We limited the inspection to five pages {\em a
priori} with the rationale that most users would not look further.  Some
projects had large numbers of results due to html documentation, repeated use
of png or svg assets, or alternate meanings of the search terms (e.g., `Visual
Studio' in every code file turning up for `visual'). Uses of GraphViz that
were documentation-only (e.g., requirements of the documentation tool or
library such as Doxygen, not to visualize dependencies) were not counted.

\begin{figure*}[!b]
    \centering
    \includegraphics[width=1.0\textwidth]{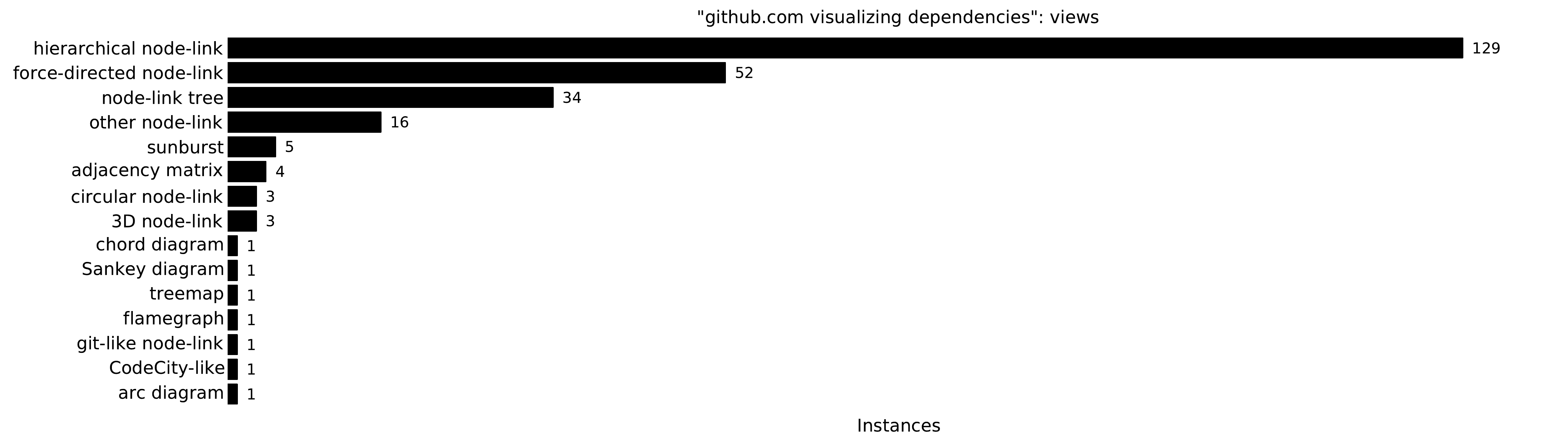}
\caption{Types of visualizations used by Github repositories.}
    \vspace{0.2in}
\label{fig:views}
\end{figure*}

Some of the links found through examining the project links led to external
sites. For those we looked at the features, gallery, and documentation pages
for evidence of dependency visualization. We did not however read the entire
documentation. Often in those cases we were unable to discern exactly how the
visualization was accomplished, so data about what libraries or tools
were used was not recorded. In our summaries, we consider the libraries used
to be `unknown' for these projects.

\begin{figure*}[!b]
    \centering
    \includegraphics[width=1.0\textwidth]{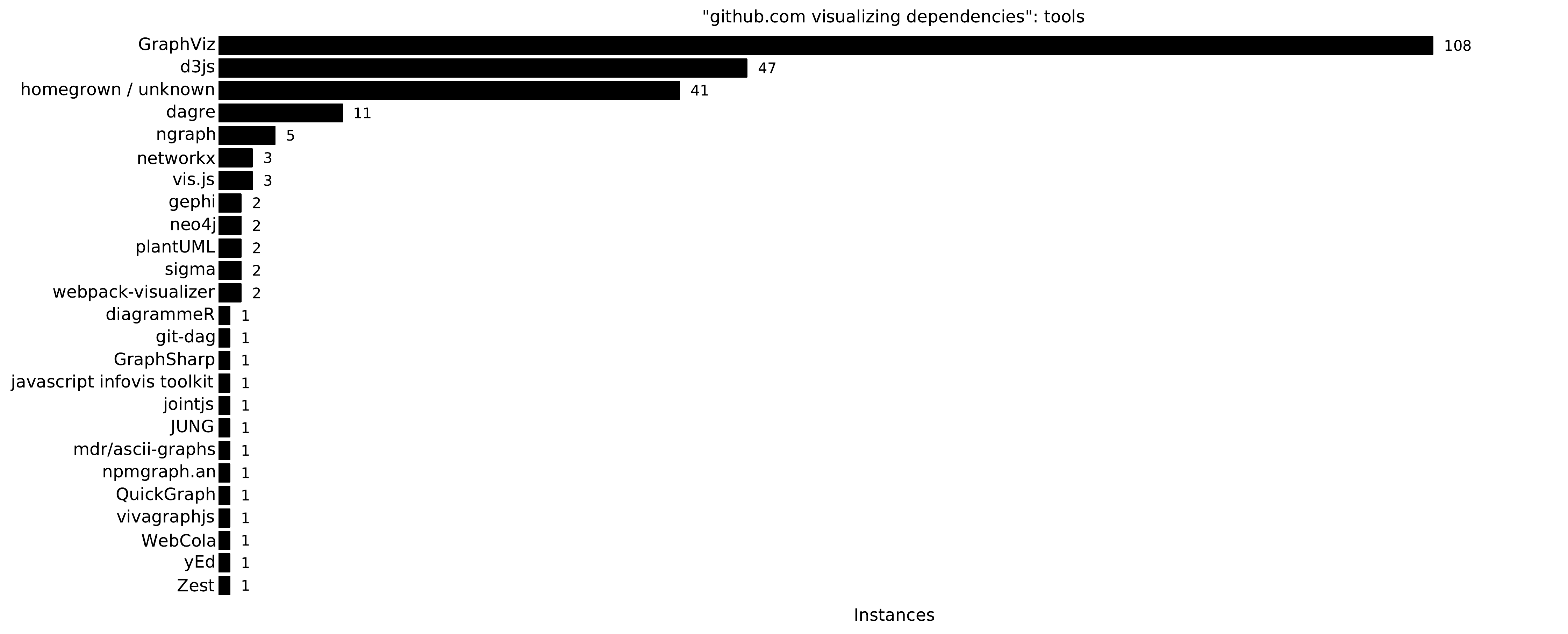}
\caption{Tools and libraries used for visualization features in Github
repositories.}
\label{fig:tools}
\end{figure*}

We found dependency visualization features in 224 of the 483 projects. Some
projects had multiple visualization options, such as outputting a \texttt{dot}
file to be rendered, rendering a png, and having a web application. Of the 224
projects, 108 had a visualization related to GraphViz through either
outputting a \texttt{dot} format file (71 projects) or the use of a
GraphViz-based rendering or layout algorithm to generate an image file, PDF,
HTML file, or application (52 projects). Eighty-three of the projects enabled
an HTML viewer for their visualization. In addition to the GraphViz-based
ones, 47 used d3js as a central tool (e.g., force-directed layout, tree,
Sankey diagram), 11 used dagre, five used ngraph, and three used visjs or
networkx. A complete list of tools can be seen in Fig.~\ref{fig:tools}. The
most common form of visualization was a layered node-link diagram (e.g.,
hierarchical layout, \texttt{dot}, Sugiyama) with 129 projects. Fifty projects
offered a force-directed layout and 34 showed a tree, 22 of which were drawn
with ASCII. 

\begin{figure*}
    \centering
    \includegraphics[width=1.0\textwidth]{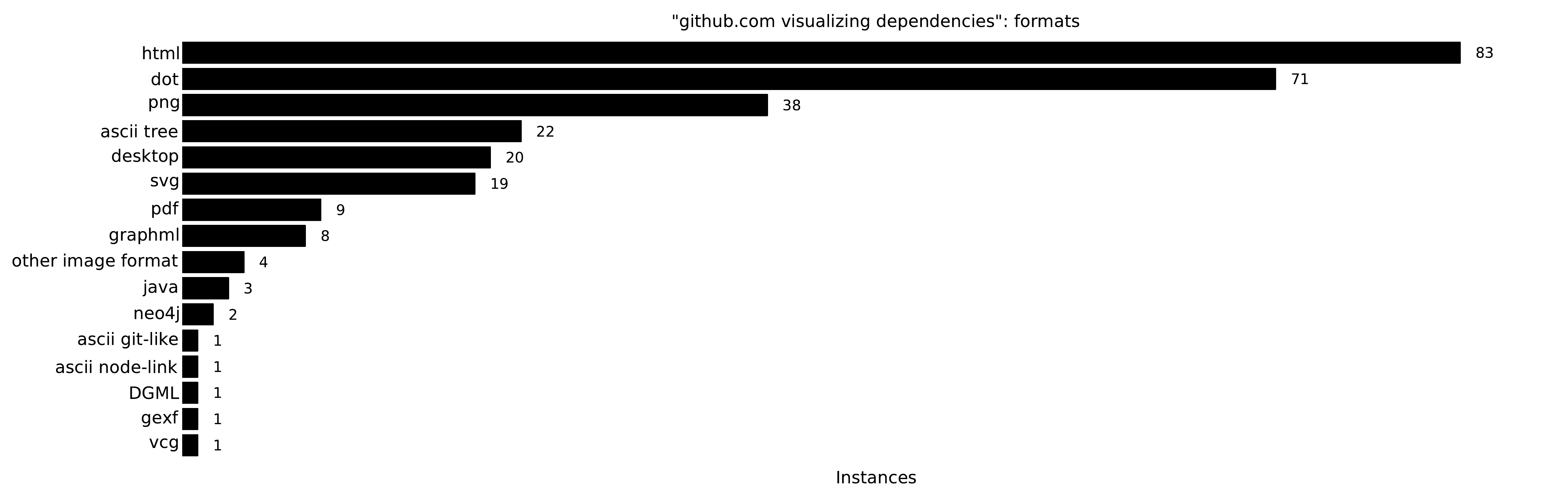}
\caption{File formats of the visualization features in Github
repositories.}
    \vspace{-0.05in}
\label{fig:formats}
\end{figure*}

Fig.~\ref{fig:views} shows the number of instances of each view we
categorized. Fig.~\ref{fig:tools} shows the number of instances of each tool or
library being used to create the visualization. Fig.~\ref{fig:formats} shows
the number of instances of each format returned by the visualization features
that we found. Web applications are uniformly categorized as `html.'

Included in these supplemental materials is a CSV listing all of the links
and their categorization, ordered by Google search ranking.


\vspace{0.1ex}

\end{document}